\documentclass[a4paper,11pt]{article}

\usepackage{latexsym}
\usepackage{amsfonts} 
\usepackage[mathscr]{euscript}
\usepackage{amsmath,mathrsfs,bm,amssymb,color}

\title{\textsf{On the semigroup generated by the renormalized Nelson Hamiltonian
}}
\date{\empty}

\author{
Tadahiro Miyao\\ 
 {\it Department of Mathematics,}
{\it Hokkaido University,}\\
{\it Sapporo 060-0810, Japan}\\
 {\tt miyao@math.sci.hokudai.ac.jp}
}

\newcommand{\one}{1}
\newcommand{\h}{\mathfrak{H}}

\newcommand{\ex}{\mathrm{e}}
\newcommand{\D}{\mathrm{dom}}

\newcommand{\Fock}{\mathfrak{F}}
\newcommand{\Ffin}{\mathfrak{F}_{\mathrm{fin}}}

\newcommand{\dG}{d\Gamma}

\newcommand{\ran}{\mathrm{ran}}

\newcommand{\la}{\langle}
\newcommand{\ra}{\rangle}

\newcommand{\BbbR}{\mathbb{R}}
\newcommand{\BbbN}{\mathbb{N}}

\newcommand{\BbbC}{\mathbb{C}}
\newcommand{\vepsilon}{\varepsilon}
\newcommand{\vphi}{\varphi}
\newcommand{\Pt}{P_{\mathrm{tot}}}
\newcommand{\Pf}{P_{\mathrm{f}}}

\newcommand{\Hf}{H_{\mathrm{f}}}

\newcommand{\Cone}{\mathfrak{P}}

\newcommand{\dm}{d}

\newcommand{\no}{\nonumber \\}

\newcommand{\Hrl}{H_{\mathrm{ren}, \Lambda}}
\newcommand{\Hr}{H_{\mathrm{ren}}}

\def\Sumoplus{\sideset{}{^{\oplus}_{n\ge 0}}\sum}

\begin{document}

\newtheorem{define}{Definition}[section]
\newtheorem{Thm}[define]{Theorem}
\newtheorem{Prop}[define]{Proposition}
\newtheorem{lemm}[define]{Lemma}
\newtheorem{rem}[define]{Remark}
\newtheorem{assum}{Condition}
\newtheorem{example}{Example}
\newtheorem{coro}[define]{Corollary}

\maketitle

\begin{abstract}
We consider the renormalized Nelson model at a fixed total momentum $P$: $\Hr(P)$;
The Hamiltonian $\Hr(P)$ is defined through an infinite energy renormalization.
We prove that $e^{-\beta \Hr(P)}$ is positivity improving 
  for all $P\in \BbbR^3$  and $\beta >0$ in the Fock representation.
  \begin{flushleft}
{\bf Mathematics Subject Classification (2010).} 
\end{flushleft}
Primary:  47A63, 47D08,  81T16;
Secondary: 47N50, 81T10
\begin{flushleft}
{\bf
Keywords. 
} 
\end{flushleft}
Nelson model;  Energy renormalization; Operator inequalities;
Positivity improving semigroups.
\end{abstract}

\section{Introduction}
In a celebrated paper \cite{Nelson}, Nelson studies the Hamiltonian, which describes the interaction of $N$
particles  with a massive Bose field. He constructs a model without the ultraviolet cutoff through an infinite 
energy renormalization.
We expect that his observation provides a hint to understand renormalization procedures
in more complicated models;
His model is nowadays called the Nelson model, and has been actively studied.
For example, Fr{\"o}hlich studies the Nelson model at a fixed total  momentum \cite{JFroehlich1,JFroehlich2}; asymptotic completeness is addressed in \cite{Ammari,DM}; existence of a ground state is proved in \cite{A. Arai,HHS,Sasaki};  functional integral
representations are constructed in \cite{GHL, LoMS,MM}, and so on \cite{AH,GW,Gross,LS,Pizzo,Spohn}.
 \medskip
  
 The cutoff Nelson Hamiltonian reads
 \begin{align}
 H_{\Lambda}=-\frac{1}{2}\Delta-g\int_{\BbbR^3} dk\frac{\chi_{\Lambda}(k)}{\sqrt{\omega(k)}}
 (e^{ik\cdot x} a(k)+e^{-ik\cdot x}a(k)^*)
 +\Hf
 \end{align}
 acting in 
 \begin{align}
 L^2(\BbbR^3) \otimes \Fock,
 \end{align}
 where $\Fock$ is the bosonic Fock space over $L^2(\BbbR^3)$. Recall that 
 \begin{align}
 \Fock=\Sumoplus L_{\mathrm{sym}}^2(\BbbR^{3n}),
 \end{align}
 where 
 $L^2_{\mathrm{sym}}(\BbbR^{3n})=\big\{
\vphi\in L^2(\BbbR^{3n})\, |\, \vphi(k_1,\dots,
k_n)=\vphi(k_{\sigma(1)}, \dots, k_{\sigma(n)})\ \mbox{a.e. }\forall
\sigma \in \mathfrak{S}_n 
\big\}
$ and $L^2_{\mathrm{sym}}(\BbbR^{0})=\BbbC$ (where $\mathfrak{S}_n$ is the permutation group on a set $\{1, 2,
\dots , n\}$).
 The single particle Schr{\"o}dinger operator $-\frac{1}{2}\Delta$ is the Hamiltonian
  of the free particle, where $\Delta$ is the $3$-dimensional Laplacian. 
 The annihilation- and creation operators of the field, $a(k)$ and $a(k)^*$,  satisfy the standard commutation relations:
 \begin{align}
 [a(k), a(k')^*]=\delta(k-k'),\ \ [a(k), a(k')]=0,\ \ k, k'\in \BbbR^3.
 \end{align}
 The field energy $\Hf$ is given by 
 \begin{align}
 \Hf=\int_{\BbbR^3} dk \omega(k)a(k)^*a(k).
 \end{align}
 The dispersion relation $\omega(k)$ is given by
 \begin{align}
 \omega(k)=\sqrt{k^2+m^2},\ \ m>0.
 \end{align}
 The ultraviolet cutoff fuction $\chi_{\Lambda}\ (\Lambda>0)$ is defined by 
 \begin{align}
 \chi_{\Lambda}(k)
 =\begin{cases}
 1, & |k| \le \Lambda\\
 0,  & |k|> \Lambda.
 \end{cases}
 \end{align}
 The prefactor $g$ is a coupling strength between the particle and the field.
 Without loss of generality, we may assume that 
 \begin{align}
 g>0.
 \end{align}
 The interaction is infinitesimally small relative to the free Hamiltonian. Hence, by the Kato-Rellich theorem,
 $H_{\Lambda}$ is self-adjoint on the domain $\D(-\Delta) \cap \D(\Hf)$ and bounded from below.
 \medskip

 The generator of translations is the total momentum operator 
 \begin{align}
 \Pt=-i \nabla+\Pf
\end{align}
with $\displaystyle 
\Pf=\int_{\BbbR^3} dk k a(k)^*a(k).
$ The total momentum is conserved, namely, $
e^{ia\cdot P_{\mathrm{tot}}}H_{\Lambda}=H_{\Lambda} e^{ia\cdot P_{\mathrm{tot}}}
$ for all $a\in \BbbR^3$. Therefore, $H_{\Lambda}$ admits the direct integral decomposition
\begin{align}
UH_{\Lambda}U^*&=\int^{\oplus}_{\BbbR^3} H_{\Lambda}(P) dP,\\
H_{\Lambda}(P)&=\frac{1}{2}(P-\Pf)^2-g\int_{\BbbR^3} dk \frac{\chi_{\Lambda}(k)}{\sqrt{\omega(k)}}
(a(k)+a(k)^*)+\Hf,
\end{align}
where $U$ is some unitary operator on $L^2(\BbbR^3) \otimes \Fock$.
$H_{\Lambda}(P)$ acts in $\Fock$.
By the Kato-Rellich theorem again, $H_{\Lambda}(P)$ is self-adjoint on $\D(\Pf^2)\cap \D(\Hf)$ and bounded from below for all $P\in \BbbR^3$. $
H_{\Lambda}(P)$ is called the cutoff Nelson Hamiltonian at a fixed total momentum $P$. 
 \medskip

 Let 
 \begin{align}
E_{\Lambda}=-g^2\int_{\BbbR^3} dk \frac{\chi_{\Lambda}(k)}{\omega(k)\{\omega(k)+k^2/2\}}. \label{DefE_k}
\end{align}
Notice that $E_{\Lambda} \to -\infty$ as $\Lambda\to \infty$.
 We define 
 \begin{align}
 \Hrl=H_{\Lambda}-E_{\Lambda},\ \ \ \ \Hrl(P)=H_{\Lambda}(P)-E_{\Lambda}.
 \end{align}
Nelson's result is stated as follows.

\begin{Thm}[Removal of UV cutoff \cite{Nelson}]
\begin{itemize}
\item[(i)] There exists a self-adjoint operator $\Hr$ bounded from below such that 
$\Hrl$ converges to $\Hr$ in strong resolvent sense as $\Lambda\to \infty$.
\item[(ii)]
For all $P\in \BbbR^3$, there exists a self-adjoint operator $\Hr(P)$ bounded from below such that 
$\Hrl(P)$ converges to $\Hr(P)$ in strong resolvent sense as $\Lambda\to \infty$.

\end{itemize}
\end{Thm}
 In this study, we are interested in the renormalized Nelson Hamiltonian at a fixed total momentum: $\Hr(P)$.
\medskip

Following Fr{\"o}hlich \cite{JFroehlich1,JFroehlich2}, we introduce a convex cone $\Fock_+$ by 
\begin{align}
\Fock_+=\Sumoplus L_{\mathrm{sym}}^2(\BbbR^{3n})_+, \label{FockCone}
\end{align}
where 
$
 L_{\mathrm{sym}}^2(\BbbR^{3n})_+=
\big\{
\vphi \in L^2_{\mathrm{sym}}(\BbbR^{3n})\, |\, \vphi(k_1,\dots, k_n)\ge
 0\ \mbox{a.e.}
\big\}$ with $
 L_{\mathrm{sym}}^2(\BbbR^{0})_+=\BbbR_+=\{r\in \BbbR\, |\, r\ge 0\} 
$. To state our results, the following terminologies are needed.
\begin{define}{\rm 
\begin{itemize}
\item A vector
$\vphi\in \Fock$ is called {\it positive} if $\vphi\in \Fock_+$;
\item
A vector 
$\vphi=\Sumoplus \vphi_n\in \Fock$ is called {\it strictly positive} if 
$\vphi_n(k_1, \dots, k_n)>0$ a.e. for all $n\in \BbbN_0=\{0\}\cup \BbbN$;
\item We say that a bounded linear operator $A$  is {\it positivity preserving} if $A$ maps $\Fock_+$
 into $\Fock_+:\ A\Fock_+\subseteq \Fock_+$;
 \item A bounded linear  operator $A$ is called {\it positivity improving} if $A\vphi$ is strictly positive whenever
  $\vphi$ is positive and $\vphi\neq 0$. $\diamondsuit$
\end{itemize}
}
\end{define}

Our main theorem
is the following.
\begin{Thm}\label{Main1}
  $e^{-\beta \Hr(P)}$ is positivity improving for all $P\in \BbbR^3$ and $\beta >0$.
\end{Thm}

The following corollary  immediately follows from   Theorems \ref{Main1} and  \ref{PFF}.

\begin{coro}
Suppose that $E(P)=\inf \mathrm{spec}(\Hr(P))$ is an eigenvalue. Then $E(P)$ is a simple eigenvalue with a strictly positive eigenvector.
\end{coro}
\begin{rem}
{\rm 
\begin{itemize}
\item[(i)] By applying methods in \cite{JFroehlich1, LMS}, we can prove that 
$E(P)$ is actually an eigenvalue, provided that $|P|<1$.
\item[(ii)]
Theorem \ref{Main1} remains true when we consider the Hamiltonian $\Hr(P)$
 with $\omega$ and $\chi_{\Lambda}$ replaced by $\omega_0(k)=|k|$ and $\chi_{\sigma}^{\Lambda}=\chi_{\Lambda}-\chi_{\sigma}$, where the infrared cutoff $\sigma$
 is chosen so that $0<\sigma <\Lambda$. 
   (Note  that when $\sigma=0$, we have to take extra care for the infrared problem, see, e.g.,  \cite{A. Arai,LoMS, Sasaki}. We will  examine such a  case  in \cite{Miyao6}.)  $\diamondsuit$
\end{itemize}
}\end{rem}

In order to explain our achievement, let us introduce the modified Nelson Hamiltonian by 
\begin{align}
H_{\varrho}(P)=\frac{1}{2} (P-\Pf)^2-g\int_{\BbbR^3} dk\frac{\varrho(k)}{\sqrt{\omega(k)}} (a(k)+a(k)^*)+\Hf,
\end{align}
where $\varrho(k)$ is real-valued.
Under the assumptions 
\begin{align}
\omega^{-1/2} \varrho,\ \  \omega^{-1} \varrho\in L^2(\BbbR^3),
\end{align}
$H_{\varrho}(P)$ is self-adjoint on $\D(\Pf^2)\cap \D(\Hf)$, and bounded from below for all $P\in \BbbR^3$.
In a famous paper \cite{JFroehlich1},  Fr{\"o}hlich has shown that, if $\varrho(k)>0$ a.e. $k$, then
$e^{-\beta H_{\varrho}(P)}$ is positivity improving for all $P\in \BbbR^3$ and $\beta >0$ in the Fock representation.
His idea has been applied to the polaron problem successfully  \cite{GeLowen,Moller2,Spohn2}; In particular, it has been proven in \cite{Miyao, Miyao3, Miyao4} that the semigroup generated by the Fr{\"o}hlich Hamiltonian without ultraviolet cutoff is positivity improving for all $P\in \BbbR^3$.
Note  that, in \cite{Sloan,Sloan2}, Sloan has proved that 
the semigroup generated by the two-dimensional polaron model without  ultraviolet cutoff
is posivitiy improving for $P=0$;  His beautiful method is different from  Fr{\"o}hlich's  approach,  and is applicable in the  Schr{\"o}dinger representation.
The primary reason for these successes is that no energy renormalization is needed, when we remove the ultraviolet cutoff from the polaron models.
\medskip

In contrast to the polaron problem, the  Hamiltonian $\Hr(P)$ is defined through an infinite energy renormalization.  By this   obstacle,     Fr{\"o}hlich's original method
only tells us that $e^{-\beta \Hr(P)}$ is postivity preserving for all $P\in \BbbR^3$ and $\beta >0$.
It has been a long standing problem to prove that $
e^{-\beta \Hr(P)}
$ is positivity improving for all $P\in \BbbR^3$.
To overcome this difficulty, we apply operator theoretic correlation inequalities studied in \cite{MS,Miyao, Miyao3, Miyao4, Miyao5}.
In our previous works on the polaron models \cite{Miyao, Miyao3, Miyao4}, we have clarified that this approach is very useful for studies
on the semigroup generated by the operator.
In the present paper, we further develop this method so that we can  get over a difficulty arising from the  infinite energy renormalization. 
\medskip

For readers' convenience, we give a brief outline of the proof of Theorem \ref{Main1} here.
For every $\kappa>0$,  let $B_{\kappa}$ be the ball of radius $\kappa$ in $\BbbR^3$ centered at the origin.
Let $\Fock^{\le \kappa}$ be the Fock space over $L^2(B_{\kappa})$ and let $\Fock^{>\kappa}$ be the Fock space over $L^2(B_{\kappa}^c)$, where $B_{\kappa}^c$ is the complement of $B_{\kappa}$.
The Fock space $\Fock$ can be factorized as 
\begin{align}
\Fock=\Fock^{\le \kappa}\otimes \Fock^{>\kappa}. \label{FockFA}
\end{align}
Corresponding to (\ref{FockFA}), $\Hr(P)$ can be decomposed as 
\begin{align}
\Hr(P)=\Hr^{\le \kappa}(P) \otimes 1\dot{+} C_{\kappa}\dot{+}1\otimes K_{\kappa}, \label{HDEC}
\end{align}
where $\dot{+}$ indicates the form sum. The local part $\Hr^{\le \kappa}(P)$ acts in $\Fock^{\le \kappa}$,
while $K_{\kappa}$ lives in $\Fock^{>\kappa}$. $C_{\kappa}$ is the cross-term.
In Section \ref{Pf1}, we will prove the following:  To show that $e^{-\beta \Hr(P)}$ improves the positivity in $\Fock$, it suffices to show that $e^{-\beta \Hr^{\le \kappa}(P)}$ improves the positivity in $\Fock^{\le \kappa}$ and $e^{-\beta K_{\kappa}}$ preserves the positivity in $\Fock^{>\kappa}$ for all $\kappa>0$.
On the other hand, we can apply  Fr{\"o}hlich's idea to see  that  $e^{-\beta \Hr^{\le \kappa}(P)}$ improves the positivity in $\Fock^{\le \kappa}$. In this way, we obtain Theorem \ref{Main1}.
The most difficult part in  the above  is the reduction of the positivity improvingness of $e^{-\beta \Hr(P)}$
 to the properties of $e^{-\beta \Hr^{\le \kappa}(P)}$.
This procedure can be achieved by extending Faris' idea in \cite{Faris} as we will see in Section \ref{Pf1}. 
\medskip

Path measure methods have been actively studied, and made remarkable progress \cite{GHL,MM,LoMS}. As far as we are aware, this methods can only cover a  case where $P=0$;
To be precise, it can be proved by a functional integral formula that $e^{-\beta \Hr(0)}$ is positivity improving in the {\it Schr{\"o}dinger representation}. Note  that this methods work for  $P=0$ only.
In contrast to this, our methods  work for all $P\in \BbbR^3$, and are effective  in the {\it Fock representation}.
  On the other hand, path measure methods can treat the Hamiltonian $\Hr+V$ with an external potential $V:\ \BbbR^3\to \BbbR$.
By using ideas in \cite{Miyao5}, our approach can also cover this case only  if  $V$ is assumed to be 
{\it ferromagnetic}\footnote{Roughly speaking, we say that $V$ is ferromagnetic if $\hat{V}(k)<0$, where $\hat{V}$ is the Frourier transformation of $V$.};
We will discuss this problem in \cite{Miyao6}.
In conclusion, our operator theoretic and path measure methods complement each other and both have specific advantages.
\medskip

Recently, Griesemer and W{\"u}nsch reported an interesting finding of  the domain property of the 
renormalized Nelson Hamiltonian in \cite{GW}.  Namely,  they showed that the domain of the Nelson model satisfies
$\D(\Hr) \cap \D(H_0)=\{0\}$, where $H_0=-\Delta+\Hf$.
Fortunately, this anomalous property unaffects our arguments in the present paper.
To be more precise, the point of our proof is the reduction of the problem to  the local properties
as we mentioned above;  this step is  essentially based on 
the algebraic relation (\ref{HDEC}), and  detailed information on the domain is unnecessary for our proof.
\medskip

The organization of the present paper is as follows:
In Section \ref{SecMono}, we briefly review some basic properties of operator theoretic correlation inequalities. 
Section \ref{2ndQuant} is devoted to study useful properties of  the second quantized operators.
In Section \ref{Pf1}, we prove Theorem \ref{Main1} by applying operator theoretic correlation inequalities.
In Appendx \ref{AppA}, we give a list of fundamental facts that are used in the main sections.

\begin{flushleft}
{\bf Acknowledgments.}

 I am  grateful to Herbert Spohn for 
comments. 
I would  like to thank an anonymous referee for helpful suggestions that
improved the present  paper. I would also like to take this opportunity to thank M. Hirokawa
and K. R. Ito for useful discussions.
This work was partially supported by   KAKENHI 18K03315.
\end{flushleft}

\section{Operator theoretic correlation inequalities}\label{SecMono}
\setcounter{equation}{0}

\subsection{Positivity preserving operators}
Let $\h$ be a complex  Hilbert space  and  let $\Cone$ be a convex cone in
$\h$.
We say that  $\Cone $ is  {\it self-dual} if 
\begin{align}
\Cone=\{x\in \h\, |\, \la x| y\ra \ge 0\  
\forall y\in \Cone
\}.
\end{align} 
Henceforth,  we  always  assume that $\Cone \neq \{0\}$.
The following properties of $\Cone$ are well-known \cite{Bos,BR1}:
\begin{Prop}\label{BasisSAC} We have  the following:
\begin{itemize}
\item[ (i)] $\Cone\cap (-\Cone)=\{0\}$.
\item[ (ii)] There exists a unique involution $j$ in $\h$ such that
                 $jx=x$ for all $x\in \Cone$.
\item[ (iii)] Each element $x\in \h$ with $jx=x$ has a unique
                 decomposition $x=x_+-x_-$,  where $x_+,x_-\in\Cone$ and
                 $\la x_+| x_-\ra=0$.
\item[ (iv)] $\h$ is linearly spanned by $\Cone$.
\end{itemize} 
\end{Prop} 
\begin{define}
{\rm 
\begin{itemize}
\item A vector $x$ is said to be  {\it positive w.r.t. $\Cone$} if $x\in
 \Cone$.  We write this as $x \ge 0$  w.r.t. $\Cone$.

 \item A vector $x \in \Cone$ is called {\it strictly positive
w.r.t. $\Cone$} whenever $\la x| y\ra>0$ for all $y\in
\Cone\backslash \{0\}$. We write this as $x>0 $
w.r.t. $\Cone$.

 \item  Let $\h_{\BbbR}=\{x\in \h\, |\, jx=x\}$, where $j$ is given in Proposition \ref{BasisSAC}.
 Let $x, y\in \h_{\BbbR}$. If $x-y\in \Cone$, then we write this as $x\ge y$ w.r.t. $\Cone$.
   $\diamondsuit$

\end{itemize} 
}
\end{define} 
\begin{example}
{\rm 
For each $d\in \BbbN$, we  set 
\begin{align}
L^2(\BbbR^d)_+=\{f\in L^2(\BbbR^d)\, |\, f(u)\ge 0\ \ \mbox{a.e. $u$}
\}.
\end{align}
$L^2(\BbbR^d)_+$ is a self-dual cone in $L^2(\BbbR^d)$.
$f\ge 0$ w.r.t. $L^2(\BbbR^d)_+$
 if and only if $f(u) \ge 0$ a.e. $u$. On the other hand, $f >0$
 w.r.t. $L^2(\BbbR^d)_+$ if and only if $f(u)>0$ a.e. $u$.
 $\diamondsuit$ 
}
\end{example} 

Let $\mathfrak{V}$ be  a dense subspace of $\h$ such that $\mathfrak{V} \cap \Cone\neq \{0\}$.\footnote{In concrete applications in Sections \ref{2ndQuant} and \ref{Pf1}, we will see that $\mathfrak{V}$ satisfies a much stronger condition: $\overline{\mathfrak{V} \cap \Cone}=\Cone$.} Set
\begin{align}
\mathscr{L}(\mathfrak{V})=\{\mbox{$A$: linear operator s.t. $\mathfrak{V}\subseteq \D(A)\cap \D(A^*),\  A\mathfrak{V} \subset \mathfrak{V},\ A^*\mathfrak{V} \subset \mathfrak{V}$}\}.
\end{align}
The following lemma is easy to check:
\begin{lemm}\label{AuxV}
We have the following:
\begin{itemize}
\item[(i)] $\mathscr{L} (\mathfrak{V})$ is a linear space.
\item[(ii)] If $A, B\in \mathscr{L}(\mathfrak{V})$, then $AB\in \mathscr{L}(\mathfrak{V})$.
\item[(iii)] If $A\in \mathscr{L}(\mathfrak{V})$, then $A^*\in \mathscr{L}(\mathfrak{V})$.
\item[(iv)] If $A\in \mathscr{L}(\mathfrak{V})$, then $\D(A) \cap \Cone 
\supseteq  \mathfrak{V}\cap \Cone
\neq \{0\}$.
\item[(v)]
If $A\in \mathscr{L}(\mathfrak{V})$, then $\D(A) \cap \h_{\BbbR} \supseteq  \mathfrak{V}\cap\h_{\BbbR} \neq \{0\}$.
\end{itemize}
\end{lemm}

\begin{define}
{\rm 
\begin{itemize}
\item Let $A\in \mathscr{L}(\mathfrak{V})$. If $A(\D(A) \cap \Cone) \subseteq \Cone$, then we write this as $A\unrhd 0$ w.r.t. $\Cone$. Remark that, by Lemma \ref{AuxV} (iv), this definition is meaningful.
In this case, we say that $A$ {\it preserves the positivity} w.r.t. $\Cone$.
\item Let $A, B\in \mathscr{L}(\mathfrak{V})$. Suppose that $A(\D(A) \cap \h_{\BbbR}) \subseteq\h_{\BbbR}$ and $B(\D(B) \cap \h_{\BbbR}) \subseteq\h_{\BbbR}$.
If $(A-B)\Big(\D(A) \cap \D(B) \cap \Cone\Big) \subseteq \Cone$, then we write this as $A\unrhd B$ w.r.t. $\Cone$. $\diamondsuit$

\end{itemize}
}
\end{define}

\begin{rem}
{\rm 
Suppose that $A$ and $B$ are bounded. Then $A\unrhd B$ w.r.t. $\Cone$ if and only if 
$\la x|Ay\ra\ge \la x|By\ra$ for all $x, y\in \Cone$. $\diamondsuit$
}
\end{rem}
\begin{example}
{\rm 
Let $F$ be a multiplication operator on $L^2(\BbbR^d)$ by the function $F(u)$.
Assume that $\|F\|_{\infty} <\infty$. If $F(u) \ge 0$ a.e., then $F\unrhd 0$ w.r.t. $L^2(\BbbR^d)_+$. $\diamondsuit$
}
\end{example}

\begin{lemm}\label{PPInq}
Let $A, A_1,  A_2,  B,  B_1, B_2\in \mathscr{L}(\mathfrak{V})$.
We have the following:
\begin{itemize}
\item[(i)]  If   $0\unlhd A$ and $0\unlhd B$ w.r.t $\Cone$, then $0\unlhd A B$ w.r.t. $\Cone$.
\item[ (ii)]  If   $0\unlhd A_1\unlhd B_1$ and $0\unlhd A_2\unlhd
 B_2$ w.r.t. $\Cone$, then $0\unlhd a A_1+b A_2 \unlhd a B_1+bB_2$ w.r.t. $\Cone$ for all
                   $a,b\in\BbbR_+$.
\item[ (iii)]  Suppose that $\Cone\cap \D(A)$ is dense in
                   $\Cone$. If $0\unlhd
                   A$ w.r.t. $\Cone$, then 
                    $0\unlhd A^*$ w.r.t. $\Cone$.
\end{itemize}
\end{lemm}
{\it Proof.} (i) and (ii) are easy to see.

(iii) Let $x\in \D(A) \cap \Cone$ and let $y\in \D(A^*) \cap \Cone$.
Then we have 
\begin{align}
 \la x|A^*y\ra=\la Ax|y\ra\ge 0. \label{xAyP}
\end{align}
 Because $\D(A) \cap \Cone$ is dense in $\Cone$, (\ref{xAyP}) holds true for all $x\in \Cone$.
 Thus, $A^*y \ge 0$, which implies that $A^*\unrhd 0$ w.r.t. $\Cone$. $\Box$
  \medskip\\
    
Let $\mathscr{B}(\h)$ be the set of all bounded linear operators on $\h$.
\begin{lemm}[\cite{Miura}]\label{Miura}
Let $A, B, C, D\in \mathscr{B}(\h)$ and let $a, b\in
 \BbbR$. 
\begin{itemize}
\item[  (i)] If $A \unrhd B \unrhd 0$ and $C\unrhd D \unrhd 0$
	     w.r.t. $\Cone$,
	     then
$AC\unrhd BD \unrhd 0$ w.r.t. $\Cone$.
\item[  (ii)] If $A \unrhd 0 $ w.r.t. $\Cone$, then $A^*\unrhd 0$ w.r.t. $\Cone$.
\end{itemize} 
\end{lemm} 
{\it Proof.} 
(i) 
By Lemma \ref{PPInq} (i), we have 
\begin{align*}
AC-BD=\underbrace{A}_{\unrhd 0}\underbrace{(C-D)}_{\unrhd 0}+
 \underbrace{(A-B)}_{\unrhd 0} \underbrace{D}_{\unrhd 0} \unrhd 0\ \ \
 \mbox{w.r.t. $\Cone$}.
\end{align*} 

(ii) follows from Lemma \ref{PPInq} (iii). $\Box$
\medskip\\

\begin{Prop}\label{WeakCl}
Let  $\mathfrak{A}=\{A\in \mathscr{B}(\h)\, |\, A\unrhd 0 \ \mbox{w.r.t. $\Cone$}\}$.
Then $\mathfrak{A}$ is a  weakly closed convex cone.
\end{Prop}
{\it Proof.} Let $\{A_n\}_{n=1}^{\infty}$ be a sequence in $\mathfrak{A}$.
Assume that $A_n$ weakly converges to $A$. Take $x, y\in \Cone$ arbitrarily.
Because $\la x|A_n y\ra\ge 0$ for all $n\in \BbbN$, we have $\la x|Ay\ra\ge 0$, which implies that $A\unrhd 0$ w.r.t. $\Cone$. 
Thus, $\mathfrak{A}$ is weakly closed.
$\Box$

\subsection{Positivity improving operators}
\begin{define}
{\rm 
Let $A\in \mathscr{B}(\h)$.
We write  $A\rhd 0$ w.r.t. $\Cone$, if  $Ax >0$ w.r.t. $\Cone$ for all $x\in
\Cone \backslash \{0\}$. 
 In this case, we say that {\it $A$ improves the
positivity w.r.t. $\Cone$.} $\diamondsuit$
}
\end{define}

The following theorem  plays  an important role.
\begin{Thm}\label{PFF}{\rm (Perron--Frobenius--Faris)}
Let $A$ be a  self-adjoint positive operator on $\h$. Suppose that 
 $0\unlhd e^{-\beta A}$ w.r.t. $\Cone$ for all $\beta \ge 0$,  and that  $\inf
 \mathrm{spec}(A)$ is an eigenvalue.
Let $P_A$ be the orthogonal projection onto the closed subspace spanned
 by  eigenvectors associated with   $\inf
 \mathrm{spec}(A)$.
 Then,  the following
 are equivalent:
\begin{itemize}
\item[ (i)] 
$\dim \ran P_A=1$ and $P_A\rhd 0$ w.r.t. $\Cone$.
\item[ (ii)] $0\lhd e^{-\beta A}$ w.r.t. $\Cone$ for all
	     $\beta >0$.
\item[ (iii)] For each $x, y\in \Cone\backslash\{0\}$,
there exists a $\beta>0$ such that $\la x| e^{-\beta A} y\ra>0$.
\end{itemize}
\end{Thm} 
{\it Proof.} See, e.g.,    \cite{Faris, Miyao,ReSi4}. $\Box$

\begin{rem}
{\rm 
(i) is equivalent to the following: The eigenvalue $\inf \mathrm{spec}(A)$ is simple with a strictly positive eigenvector. $\diamondsuit$
}
\end{rem}
\section{Second quantized operators}\label{2ndQuant}
\setcounter{equation}{0}

We briefly summarize necessary
results concerning the second quantized operators. 
As to basic definitions, we refer to \cite{BraRob2} as an accessible text.

\subsection{Basic definitions}
Let $\h$ be a complex Hilbert space.
The bosonic  Fock space over $\h$ is defined by 
\begin{align}
\Fock(\h)=\Sumoplus 
\Fock^{(n)}(\h),
\ \ \ \Fock^{(n)}(\h)=\h^{\otimes_{\mathrm{s}}n},
\end{align} 
where $\h^{\otimes_{\mathrm{s}}n}$  is the $n$-fold
symmetric tensor product of $\h$ with convention 
$\h^{\otimes_{\mathrm{s}}0}=\BbbC$. 
$\Fock^{(n)}(\h)$ is called the {\it $n$-boson subspace}.
A finite particle subspace $\Ffin(\h)$ is defined by 
\begin{align}
\Ffin(\h)=\bigg\{\vphi=\Sumoplus \vphi_n \in \Fock(\h)\, \bigg|\, \mbox{$\exists N\in \BbbN_0$ such that $\vphi_n=0$ for all $n\ge N$}\bigg\}.
\end{align}

We denote by $a(f)\, (f\in \h)$ the annihilation operator on
$\Fock(\h)$, its adjoint $a(f)^*$, called the creation operator, is defined by
\begin{align}
a(f)^*\vphi=\sideset{}{^{\oplus}_{n\ge 1}}\sum
 \sqrt{n}S_n (f\otimes \vphi_{n-1})\label{DefCrea}
\end{align} 
for $\vphi=\sum_{n\ge 0}^{\oplus} \vphi_n \in \D(a(f)^*)$, where
$S_n$ is the symmetrizer  on $\Fock^{(n)}(\h)$.
The annihilation- and creation operators satisfy
the cannonical commutation relations (CCRs)
\begin{align}
[a(f), a(g)^*]=\la f| g\ra, \ \ [a(f), a(g)]=0=[a(f)^*, a(g)^*]
\end{align} 
on $\Ffin(\h)$.

Let $C$ be a contraction operator on $\h$, that is ,
$\|C\|\le 1$. Then we define a contraction operator $\Gamma(C)$ on
$\Fock(\h)$
by 
\begin{align}
\Gamma(C)=\Sumoplus C^{\otimes n}
\end{align} 
with $C^{\otimes 0}=\one$, the identity operator. 

For a self-adjoint operator $A$ on $\h$, let us introduce 
\begin{align}
\dG(A)=0\oplus \sideset{}{^{\oplus}_{n\ge 1}}\sum
 \sideset{}{_{n\ge k\ge 1}}\sum
\one^{\otimes (k-1)}\otimes A \otimes \one^{\otimes (n-k)}  
\end{align} 
acting in   $\Fock(\h)$. Then  $\dG(A)$ is essentially self-adjoint. We denote its
closure by the same symbol. 

If $A$ is positive, then one has 
\begin{align}
\Gamma(e^{-tA})=e^{-t \dG(A)},\ \ \ t \ge 0. \label{SemiP2}
\end{align} 

The following proposition is well-known.
\begin{Prop}
Let $A$ be a positive self-adjoint operator. 
For each $f\in \D(A^{-1/2})$, 
we have the following operator inequalities:
\begin{align}
a(f)^*a(f)&\le \|A^{-1/2}f\|^2 (\dG(A)+\one),\label{CreAnnInq}\\
a(f)a(f)^*&\le \|A^{-1/2}f\|^2 (\dG(A)+\one),\label{CreAnnInq2}\\
\dG(A)+a(f)+a(f)^*&\ge -\|A^{-1/2}f\|^2. \label{VHove}
\end{align} 
\end{Prop}

\subsection{Fock space over $L^2(\BbbR^3)$}
In this study, the bosonic Fock space over $L^2(\BbbR^3)$  is important. We simply write  it  as 
\begin{align}
\Fock=\Fock(L^2(\BbbR^3)).
\end{align} 
The $n$-boson subspace $\Fock^{(n)}=L^2(\BbbR^3)^{\otimes_{\mathrm{s}}n}$ is naturally
identified with $
L^2_{\mathrm{sym}}(\BbbR^{3n})
$.
Hence 
\begin{align}
\Fock=\BbbC \oplus \sideset{}{^{\oplus}_{n\ge 1}}\sum
 L^2_{\mathrm{sym}}(\BbbR^{3n}). \label{FockId}
\end{align}
The annihilation- and creation operators are symbolically expressed as 
\begin{align}
a(f)=\int_{\BbbR^3}\dm k\, \overline{f(k)}a(k),\ \
 a(f)^*=\int_{\BbbR^3}\dm k\, f(k) a(k)^*.
\end{align}
If $F$ is a multipilication operator by the function $F(k)$,
then $\dG(F)$ is formally written as 
\begin{align}
\dG(F)=\int_{\BbbR^3}\dm k\, F(k)a(k)^* a(k).
\end{align}  
Note  that $\dG(F) \restriction L^2_{\mathrm{sym}}(\BbbR^{3n})$ is a mutiplication operator by the function $F(k_1)+\cdots+F(k_n)$.

\subsection{The Fr\"ohlich cone}
Let $\Fock_+$ be a convex cone defined by (\ref{FockCone}). We begin with the following lemma:
\begin{lemm}
$\Fock_+$
 is a self-dual cone in 
$\Fock$. 
 \end{lemm}
{\it Proof.}
It suffices to show that $L^2_{\mathrm{sym}}(\BbbR^{3n})_+$ is a self-dual cone for all $n\in\BbbN_0$.
To this end, we set $\Cone=L^2_{\mathrm{sym}}(\BbbR^{3n})_+$. It is easy to check that $\Cone \subseteq \Cone^{\dagger}$.
To prove the converse, we note  the following fact:
 Let $\psi\in L^2_{\mathrm{sym}}(\BbbR^{3n})$.  $\psi\ge 0$ w.r.t. $L^2_{\mathrm{sym}}(\BbbR^{3n})_+$ if and only if 
 \begin{align}
 \la f_1\otimes \cdots \otimes f_n|\psi\ra \ge 0\label{PPEqui}
 \end{align}
    for all $f_1, \dots, f_n\in L^2(\BbbR^{3})_+$, 
 where $
 (
 f_1\otimes \cdots \otimes f_n
 )(k_1, \dots, k_n)=f_1(k_1)\cdots f_n(k_n)
 $. But it is easy to prove  (\ref{PPEqui})  for each $\psi\in \Cone^{\dagger}$. $\Box$

\begin{define}[\cite{JFroehlich1,JFroehlich2}]{\rm
The self-dual cone $\Fock_+$ is called the {\it Fr\"ohlich cone}. $\diamondsuit$
}
\end{define}

\begin{lemm}
We have the following:
\begin{itemize}
\item[(i)] $a(f)$ and $a(f)^*\in \mathscr{L}(\Ffin)$ for all $f\in L^2(\BbbR^3)$.
\item[(ii)] If $F$ is a multiplication operator such that $\|F\|_{\infty} \le 1$, then $\Gamma(F) \in \mathscr{L}(\Ffin)$.
\end{itemize}
\end{lemm}
By using  the above lemma, we can 
discuss operator inequalities given in  Section \ref{SecMono}.

The following lemma will be useful.
\begin{lemm}\label{ClPos}
Let $\Fock_{\mathrm{fin}, +}=\Ffin \cap \Fock_+$.  Then $
\overline{\Fock_{\mathrm{fin}, +}}=\Fock_+
$, where the bar indicates the closure in the strong topology.
\end{lemm}

We summarize  properties of operators on $\Fock$ below.
All propositions were proven in  \cite{Miyao3}. For reader's convenience, we will provide proofs.
\begin{Prop}
\label{PPFockI}
 Let $C$ be a contraction  operator on $L^2(\BbbR^3)$.  If
$C\unrhd 0$ w.r.t. $L^2(\BbbR^3)_+$,   then we have 
$\Gamma(C)\unrhd 0$ w.r.t. $\Fock_+$.
 \end{Prop}
 {\it Proof.}
 Let $f_1, \dots, f_n\in L^2(\BbbR^3)_+$.
 Because $C\unrhd 0$ w.r.t. $L^2(\BbbR^3)_+$, we have $Cf_j \ge 0$ w.r.t. $L^2(\BbbR^3)_+$, which implies that $Cf_1\otimes \cdots \otimes Cf_n \ge 0$ w.r.t. $L^2(\BbbR^{3n})_+$.
 Thus,
 \begin{align}
 \la f_1\otimes \cdots \otimes f_n|C^{\otimes n}\psi\ra=
 \la C f_1\otimes \cdots \otimes C f_n|\psi\ra \ge 0
 \end{align}
  for all $f_1, \dots, f_n\in L^2(\BbbR^{3})_+$, which implies that $C^{\otimes n} \unrhd 0$
 w.r.t. $L^2_{\mathrm{sym}}(\BbbR^{3n})_+$. $\Box$

 \begin{Prop}\label{PPFock3}
 Let $B$ be  a positive  self-adjoint operator. If $e^{- t B}\unrhd 0$
w.r.t. $L^2(\BbbR^3)_+$ for all $t\ge 0$, then  $e^{-t \dG(B)}\unrhd 0$
w.r.t. $\Fock_+$ for all $t\ge 0$.
\end{Prop} 
{\it Proof.} By (\ref{SemiP2}) and Proposition \ref{PPFockI}, we obtain the desired assertion. $\Box$

\begin{Prop}\label{PPFockII}
If $f \ge 0$ w.r.t. $L^2(\BbbR^3)_+$, then $a(f)^*\unrhd 0$ and
 $a(f)\unrhd 0$ w.r.t. $\Fock_+$.
\end{Prop} 
{\it Proof}.  Let $\vphi=\Sumoplus \vphi_n\in \Fock_+\cap \D(a(f)^*)$. By (\ref{DefCrea}), 
we have
\begin{align}
\big(
a(f)^* \vphi
\big)_{n+1} (k_1, \dots, k_{n+1})= \frac{1}{\sqrt{n+1}} \sum_{j=1}^{n+1} \underbrace{f(k_j)}_{\ge 0} 
\underbrace{\vphi_n(k_1, \dots, \hat{k}_j, \dots, k_{n+1})}_{\ge 0} \ge 0,
\end{align}
where $\hat{k}_j$ indicates the omission of $k_j$. Thus, $a(f)^*\unrhd 0$ w.r.t. $\Fock_+$.
Because $a(f)=(a(f)^*)^*$, we have $a(f) \unrhd 0$ w.r.t. $\Fock_+$ by Lemmas  \ref{PPInq} (iii)  and \ref{ClPos}. $\Box$

\begin{Prop}\label{ErgoFock}{\rm (Ergodicity)}
For each $f\in L^2(\BbbR^3)$, let $\phi(f)$ be a  linear
 operator
 defined by
\begin{align}
\phi(f)=a(f)+a(f)^*.
\end{align} 
Note that $\phi(f)$ is essentially self-adjoint. We denote its closure by the same symbol.
If $f >0$ w.r.t. $L^2(\BbbR^3)_+$, that is, $f(k)>0$ a.e. $k$, then
$\phi(f)$ is ergodic in the sense  that, for any $\vphi, \psi\in
 \Fock_{\mathrm{fin}, +}\backslash \{0\}$, there exists an $n\in \BbbN_0$ such
 that $\la \vphi|\phi(f)^n \psi\ra >0$.
\end{Prop} 
{\it Proof.}  Choose $\vphi, \psi\in \Fock_{\mathrm{fin}, +}\backslash \{0\}$, arbitrarily.
We can express $\vphi$ and $\psi$ as 
\begin{align}
 \vphi=\Sumoplus \vphi_n,\ \ \ \psi=\Sumoplus \psi_n.
\end{align}
Because $\vphi$ and $\psi$ are non-zero, there exist $p, q\in \BbbN_0$ such that $\vphi_p \neq 0$ 
and $\psi_q\neq 0$.
Under the identifications
\begin{align}
\vphi_p=\Sumoplus \delta_{np} \vphi_n,\ \ \ \psi_q=\Sumoplus \delta_{nq}\psi_n, \label{FinIdn}
\end{align}
we have $\vphi \ge \vphi_p$ and $\psi\ge \psi_q$ w.r.t. $\Fock_+$, where $\delta_{mn}$ is the Kronecker delta.
By Proposition \ref{PPFockII}, we have 
\begin{align}
\la \vphi|\phi(f)^{p+q} \psi\ra \ge  \la \vphi_p|\phi(f)^{p+q} \psi_q\ra. \label{ErgPf1}
\end{align}
Because $\phi(f)^p\unrhd a(f)^p$ and $\phi(f)^q \unrhd a(f)^q$ w.r.t. $\Fock_+$, we have 
\begin{align}
\mbox{the RHS of (\ref{ErgPf1})} \ge \la a(f)^p \vphi_p| a(f)^q \psi_q\ra.\label{ErgPf2}
\end{align}
Remark that 
\begin{align}
a(f)^p\vphi_p=\sqrt{p!} \la f^{\otimes p}|\vphi_p\ra\Omega,\ \ \ 
a(f)^q\psi_q=\sqrt{q!} \la f^{\otimes q}|\psi_q\ra\Omega,
\end{align}
where $\Omega=1\oplus 0\oplus0\oplus \cdots$ is the Fock vacuum. 
Since $
\la f^{\otimes p}|\vphi_p\ra>0
$ and 
$
\la f^{\otimes q}|\psi_q\ra>0
$, we get, by (\ref{ErgPf1}) and (\ref{ErgPf2}),
\begin{align}
\la \vphi|\phi(f)^{p+q} \psi\ra \ge  \sqrt{p! q!} \la f^{\otimes p}|\vphi_p\ra  \la f^{\otimes q}|\psi_q\ra>0.
\end{align}
Thus we are done. $\Box$

\subsection{Local properties}
Let $B_{\kappa}$ be a ball of radius $\kappa$ in $\BbbR^3$ centered at the origin and let
$\chi_{\kappa}$ be a function on $\BbbR^3$ defined by
$\chi_{\kappa}(k)=1$ if $k \in B_{\kappa}$ and $\chi_{\kappa}(k)=0$
otherwise. 
Then as a multiplication operator, $\chi_{\kappa}$
is an orthogonal projection on $L^2(\BbbR^3)$ and
 $Q_{\kappa}=\Gamma(\chi_{\kappa})$ is  an orthogonal projection  on $\Fock$ as well.
 We remark the following properties:
\begin{itemize}
 \item If $\kappa_1 \ge \kappa_2$, then 
 $Q_{\kappa_1} \ge Q_{\kappa_2}$.
 \item $Q_{\kappa}$ strongly converges to $1$ as $\kappa\to \infty$.
 \end{itemize}
Let us define the local Fock space by
\begin{align}
\Fock^{\le \kappa}=Q_{\kappa}\Fock.
\end{align} 
 Since $\chi_{\kappa}L^2(\BbbR^3)=L^2(B_{\kappa})$, $\Fock^{\le \kappa}$ can be
identified with $\Fock(L^2(B_{\kappa}))$. 
In what follows, $\Ffin^{\le \kappa}$ denotes  $\Ffin (L^2(B_{\kappa}))$.
The following fact will be useful:
\begin{align}
\Fock=\overline{\bigcup_{\kappa\ge 0} \Fock^{\le \kappa}}.
\end{align}

\begin{Prop}\label{QPP}
For each $\kappa \ge 0$, we set
 $Q_{\kappa}^{\perp}=\one-Q_{\kappa}$. Then we have  the following:
\begin{itemize}
\item[  (i)] $Q_{\kappa}\unrhd 0$ w.r.t. $\Fock_+$.
\item[ (ii)] $Q_{\kappa}^{\perp} \unrhd 0$ w.r.t. $\Fock_+$.
\end{itemize} 
\end{Prop} 
{\it Proof.} (i) immediately follows from Proposition \ref{PPFockI}.

(ii) Under the identification (\ref{FockId}), we see
\begin{align}
(Q_{\kappa}\vphi_n)(k_1, \dots, k_n)=\Bigg[ \prod_{j=1}^n\chi_{\kappa}(k_j)\Bigg]
 \vphi_n(k_1,\dots, k_n)
\end{align} 
for each $\vphi_n\in L^2_{\mathrm{sym}}(\BbbR^{3n})$. Hence 
\begin{align}
(Q^{\perp}_{\kappa}\vphi_n)(k_1,\dots, k_n)=\Bigg\{1-\prod_{j=1}^n
 \chi_{\kappa}(k_j)\Bigg\} \vphi_n(k_1,\dots, k_n). \label{OrthQ}
\end{align} 
If $\vphi_n(k_1,\dots, k_n)\ge 0$ a.e., then the right hand side of
(\ref{OrthQ}) is positive for a.e. $k_1, \dots, k_n$  because
$1-\prod_{j=1}^n \chi_{\kappa}(k_j)\ge 0$. This means that 
$Q^{\perp}_{\kappa} \unrhd 0$ w.r.t. $\Fock_{+}$. $\Box$
\medskip\\

We remark the following:
\begin{align}
a(f)Q_{\kappa}&=a(\chi_{\kappa}f)=\int_{|k|\le \kappa}\dm k\,
 \overline{f(k)} a(k),\\
Q_{\kappa} a(f)^*&=a(\chi_{\kappa}f)^*=\int_{|k|\le \kappa}\dm k\,
 f(k) a(k)^*,\\
\dG(F) Q_{\kappa}&=\dG(\chi_{\kappa}F)=\int_{|k|\le \kappa}\dm k\, F(k)a(k)^*a(k).
\end{align}
By these facts, we obtain the following proposition.
\begin{Prop}\label{CommuL}
We have the following:
\begin{itemize}
\item[(i)] $[Q_{\kappa}, a(f)]=Q_{\kappa} a((1-\chi_{\kappa})f)$ on $\Ffin$.
\item[(ii)] $[Q_{\kappa}, \dG(F)]=0$  on $\D(\dG(F))$.
\end{itemize}
\end{Prop}

Next let us introduce  a  natural self-dual cone in $\Fock^{\le \kappa}$.
To this end, define
\begin{align}
\Fock^{\le \kappa}_{n, +}=\big\{
\vphi\in L^2_{\mathrm{sym}}(B_{\kappa}^{\times n})\, |\,
 \vphi(k_1,\dots, k_n) \ge 0\ \   a.e.
\big\}
\end{align}
with $\Fock_{0, +}^{\le \kappa}=\BbbR_+$.  Each $\Fock_{
n, +}^{\le \kappa}$ is a self-dual cone in
$L^2(B_{\kappa})^{\otimes_{\mathrm{s}}n}=L^2_{\mathrm{sym}}(B_{\kappa}^{\times
n})$.
\begin{define}
{\rm
The {\it  local Fr\"ohlich cone } is defined by 
\begin{align}
\Fock_{+}^{\le \kappa}=\Sumoplus \Fock_{n,  +}^{\le \kappa}.
\end{align} 
$\Fock_{+}^{\le \kappa}$ is a self-dual cone in $\Fock^{\le \kappa}$. 
As before, we define $\Fock_{\mathrm{fin}, +} ^{\le \kappa}=\Ffin^{\le \kappa } \cap \Fock_+^{\le \kappa}$.
Note  that $
\overline{\Fock_{\mathrm{fin}, +} ^{\le \kappa}}=\Fock_+^{\le \kappa}
$.
$\diamondsuit$
}
\end{define} 

\begin{Prop}\label{LocalPropErgo}
Propositions \ref{PPFockI}, \ref{PPFock3}, \ref{PPFockII} and \ref{ErgoFock} are still
 true even if one replaces $L^2(\BbbR^3)_+$, $\Fock_+$  and $\Fock_{\mathrm{fin}, +}$by
 $L^2(B_{\kappa})_+$, $\Fock_{ +}^{\le \kappa}$ and $\Fock_{\mathrm{fin}, +} ^{\le \kappa}$,  respectively. 
\end{Prop}

\subsection{Decomposition properties}
Let $\mathfrak{h}_1$ and $\mathfrak{h}_2$ be complex Hilbert spaces.
Remark the following 
factorization property:
\begin{align}
\Fock(\mathfrak{h}_1\oplus \mathfrak{h}_2)=\Fock(\mathfrak{h}_1)\otimes \Fock(\mathfrak{h}_2)
. \label{GFact}
\end{align}
Corresponding to this, we have the following:

\begin{itemize}
\item For each 
$
  f\in \mathfrak{h}_1,\ 
g\in \mathfrak{h}_2
$,
\begin{align}
a(f\oplus g)=a(f)\otimes \one +\one \otimes a(g). \label{Fac1}
\end{align}

\item 
Let $A$ and  $B$ be self-adjoint operators. We have 
\begin{align}
\dG(A\oplus B)=\{\dG(A)\otimes \one+\one \otimes \dG(B)\}^-,\label{Fac2}
\end{align}
  where $\{\cdots\}^-$ indicates the closure of $\{\cdots\}$.
  \item
  Let $C$ and $D$ be contraction operators. We have
  \begin{align}
  \Gamma(C\oplus D)=\Gamma(C)\otimes \Gamma(D).
  \end{align}
\end{itemize}
For each $\kappa >0$, we have the following identification:
\begin{align}
L^2(\BbbR^3)=L^2(B_{\kappa}) \oplus L^2(B_{\kappa}^c), \label{L2Dec}
\end{align}
where  $B_{\kappa}^c$ indicates the complement of $B_{\kappa}$. 
Using (\ref{GFact}) and (\ref{L2Dec}), we have
\begin{align}
\Fock=&\Fock^{\le \kappa}\otimes \Fock^{>\kappa}, \label{FactFock}
\end{align}
where $\Fock^{>\kappa}=\Fock(L^2(B_{\kappa}^c))$. 
Thus, we have 
\begin{align}
\Fock=&\Sumoplus \Fock^{\le \kappa}\otimes L^2_{\mathrm{sym}}((B_{\kappa}^c)^{\times n})\no
=&\Fock^{\le \kappa} \oplus
\bigg[
\sideset{}{^{\oplus}_{n\ge 1}}\sum \Fock^{\le \kappa}\otimes L^2_{\mathrm{sym}}((B_{\kappa}^c)^{\times n})
\bigg], \label{Iden1}
\end{align}
where 
 $
 L^2_{\mathrm{sym}}((B_{\kappa}^c)^{\times 0}):=\BbbC
$.
The following lemma will be useful.
\begin{lemm}\label{QAct}
Let $\psi =\Sumoplus \psi_n(k_1, \dots, k_n)\in \Fock$.
For each $\kappa>0$, we have 
\begin{align}
Q_{\kappa} \psi=\psi_{\kappa}\otimes \Omega^{>\kappa}, \label{ProjQ}
\end{align}
  where
$\Omega^{>\kappa}$ is the Fock vacuum in $\Fock^{>\kappa}$ and 
\begin{align}
\psi_{\kappa}=\Sumoplus \Bigg[\prod_{\ell=1}^n \chi_{\kappa}(k_{\ell})\Bigg] \psi_n(k_1, \dots, k_n). \label{Defpsik}
\end{align}
\end{lemm}

A natural self-dual cone in $\Fock^{>\kappa}$ is given by 
\begin{align}
\Fock_+^{>\kappa}=\Sumoplus L^2_{\mathrm{sym}}((B_{\kappa}^c)^{\times n})_+,
\end{align}
where  $
 L^2_{\mathrm{sym}}((B_{\kappa}^c)^{\times 0})_+:=\BbbR_+
$.  As before, we set $\Ffin^{>\kappa}=\Ffin(L^2(B_{\kappa}^c))$
 and $\Fock_{\mathrm{fin}, +}^{>\kappa}=\Ffin^{>\kappa} \cap \Fock_+^{>\kappa}$.

\begin{Prop}
Propositions \ref{PPFockI}, \ref{PPFock3}, \ref{PPFockII} and \ref{ErgoFock} are still
 true even if one replaces $L^2(\BbbR^3)_+$, $\Fock_+$  and $\Fock_{\mathrm{fin}, +}$by
 $L^2(B_{\kappa}^c)_+$, $\Fock_{ +}^{> \kappa}$ and $\Fock_{\mathrm{fin}, +} ^{> \kappa}$,  respectively. 
\end{Prop}

The self-dual cone $\Fock_+$ can be expressed as 
\begin{align}
\Fock_+=\Fock^{\le \kappa}_+ \oplus
\bigg[\sideset{}{^{\oplus}_{n\ge 1}}\sum \Fock^{\le \kappa}_+\otimes L^2_{\mathrm{sym}}((B_{\kappa}^c)^{\times n})_+\bigg], \label{DSumSC}
\end{align}
where  \begin{align}
\Fock^{\le \kappa}_+\otimes L^2_{\mathrm{sym}}((B_{\kappa}^c)^{\times n})_+
=\Big\{\psi\in \Fock^{\le \kappa}\otimes L^2_{\mathrm{sym}}((B_{\kappa}^c)^{\times n})\, \Big|\, 
\psi(k_1, \dots, k_n) \ge0\ \mbox{w.r.t. $\Fock_+^{\le \kappa}$ a.e.}\Big\}.
\end{align}

\begin{Thm}\label{PCup}
We have the following:
\begin{itemize}
\item[ (i)] $Q_{\kappa}\Fock_+=\Fock^{\le \kappa}_+$.
\item[ (ii)]
$\displaystyle 
\Fock_+=\overline{\bigcup_{\kappa >0} \Fock_+^{\le \kappa}}
$.
\end{itemize}
\end{Thm}
{\it Proof.} 
(i) This immediately follows from (\ref{DSumSC}).

(ii)
With the identification $
\Fock_+^{\le \kappa} = \Fock_+^{\le \kappa} \oplus \{0\}
$, we know that $\Fock_+\supseteq 
\Fock_+^{\le \kappa} 
$ by (\ref{DSumSC}). Hence, $
\displaystyle 
\Fock_+\supseteq  \overline{\bigcup_{\kappa>0} \Fock_+^{\le \kappa}}
$. 

Let $\psi\in \Fock_+$. 
For each $\kappa>0$, we know that $Q_{\kappa} \psi\in \Fock_+^{\le \kappa}$ by (\ref{ProjQ}).
Because $Q_{\kappa}$ strongly converges  to $\one$ as $\kappa\to \infty$, we conclude that 
$\psi \in \overline{\bigcup_{\kappa>0} \Fock_+^{\le \kappa}}$. $\Box$

\begin{lemm}\label{TensPP}
Let $\psi\in \Fock$. The following (i) and (ii) are equivalent:
\begin{itemize}
\item[(i)] $\psi\ge 0$ w.r.t. $\Fock_+$.
\item[(ii)] $\la \xi \otimes \eta|\psi\ra\ge 0$ for all $\xi\in \Fock_+^{\le\kappa}$ and $\eta \in \Fock_+^{>\kappa}.$
\end{itemize}

\end{lemm}
{\it Proof.}
(ii) $\Longrightarrow$ (i):
Without loss of generality, we may assume that $\psi\in \Ffin$.
Thus, it suffices to consider the case where $\psi=\psi_1\otimes \psi_2$ with $\psi_1\in \Ffin^{\le \kappa}$
and $\psi_2\in \Ffin^{>\kappa}$.
Because $
\la \xi\otimes \eta|\psi\ra=\la \xi|\psi_1\ra\la \eta|\psi_2\ra \ge 0
$, we can choose $\psi_1$ and  $\psi_2$ such that $\psi_1\ge 0$ w.r.t. $\Fock_+^{\le \kappa}$ and $\psi_2\in \Fock_+^{>\kappa}$.
Thus, we conclude that $\psi\ge 0$ w.r.t. $\Fock_+$.

(i) $\Longrightarrow$ (ii): By arguments similar to those in the above, it suffices to consider the case $
\psi=\psi_1\otimes \psi_2
$ with $\psi_1\in \Fock_+^{\le \kappa}$ and $\psi_2\in \Fock_+^{>\kappa}$.
In this case,  we easily check that $\la \xi \otimes \eta|\psi\ra\ge 0$ for all $\xi\in \Fock_+^{\le\kappa}$ and $\eta \in \Fock_+^{>\kappa}.$ $\Box$

\begin{Prop}\label{TensPPAB}
Let $A\in \mathscr{B}(\Fock^{\le \kappa})$ and $B\in \mathscr{B}(\Fock^{> \kappa})$.
If $A\unrhd 0$ w.r.t.  $\Fock_+^{\le \kappa}$ and $B\unrhd 0$ w.r.t. $\Fock_+^{>\kappa}$, then
$A\otimes B\unrhd 0$ w.r.t. $\Fock_+$.
\end{Prop}
{\it Proof.}
Let $\xi\in \Fock_+^{\le \kappa}$ and let $\eta\in \Fock_+^{>\kappa}$.
By the assumption, we have $A^*\xi\ge 0$ and $B^*\eta\ge 0$.
Thus, by Lemma \ref{TensPP}, 
\begin{align}
\la \xi \otimes \eta|A\otimes B\psi\ra=\la (A^* \xi)\otimes  (B^*\eta)| \psi\ra \ge 0.
\end{align}
By Lemma \ref{TensPP} again, we have $A\otimes B\psi\ge 0$ w.r.t. $\Fock_+$. $\Box$

\section{Proof of Theorem \ref{Main1}}\label{Pf1}
\setcounter{equation}{0}
\subsection{Decomposition of $\Hrl(P)$}

 In what follows, we always assume that $\kappa <\Lambda$.
Let $F$ be a  real-valued measurable function on $\BbbR^3$. Suppose that $F(k)$ is finite for almost everywhere. Then $\dG(F)$ is essentially self-adjoint. 
For each $\kappa>0$, we set $F^{\le \kappa}=\chi_{\kappa} F$ and $F^{>\kappa}=(1-\chi_{\kappa}) F$.
By (\ref{Fac2}) and (\ref{FactFock}), we have 
\begin{align}
\dG(F)=\{\dG(F^{\le \kappa}) \otimes \one +\one \otimes \dG(F^{>\kappa})\}^{-}.
\end{align}
Keeping this fact in mind, we set 
\begin{align}
P_{\mathrm{f}, j}^{\le \kappa}=\dG(k_j \chi_{\kappa}),\ \ P_{\mathrm{f}, j}^{> \kappa}=\dG(k_j (1-\chi_{\kappa})),\ \ j=1,2,3.
\end{align}
Remark the following formulas:
\begin{align}
\dG(\omega)&=\dG(\omega^{\le \kappa}) \otimes 1+1\otimes \dG(\omega^{>\kappa}), \label{Dec1}\\
P_{\mathrm{f}, j}
&=\{P_{\mathrm{f}, j}^{\le \kappa}\otimes \one +\one \otimes  P_{\mathrm{f}, j}^{> \kappa}\}^-,\label{Dec2}\\
a(f)&= a(\chi_{\kappa} f)\otimes 1+1\otimes a((1-\chi_{\kappa})f).\label{Dec3}
\end{align}

Let 
\begin{align}
E^{\Lambda}_{\kappa}
=-g^2\int_{\BbbR^3} dk \frac{\chi_{\kappa}^{\Lambda}(k)}{\omega(k)\{\omega(k)+k^2/2\}},\ \ \
\chi_{\kappa}^{\Lambda}=\chi_{\Lambda}-\chi_{\kappa}.
\end{align}
Note that $E_{\kappa}^{\Lambda}=E_{\Lambda}-E_{\kappa}$, where $E_{\Lambda}$ is 
 defined by (\ref{DefE_k}), while $E_{\kappa}$ is defined by  (\ref{DefE_k}) with $\Lambda$ replaced by $\kappa$.
  Using (\ref{Dec1}), (\ref{Dec2}) and (\ref{Dec3}), we have
\begin{align}
\Hrl(P)
=H_{\mathrm{ren}}^{\le  \kappa}(P)\otimes\one +\one \otimes   K_{\kappa, \Lambda}
-(P-\Pf^{\le \kappa}) \cdot \Pf^{>\kappa}
,
\end{align}
where 
\begin{align}
\Hr^{\le \kappa}(P)=&\frac{1}{2}(P-\Pf^{\le\kappa})^2-g
\int_{\BbbR^3}dk\frac{\chi_{\kappa}(k)}{\sqrt{\omega(k)}}
(a(k)+a(k)^*)
+\dG(\omega^{\le \kappa})
-E_{\kappa},\\
K_{\kappa, \Lambda}
=&
\frac{1}{2}(\Pf^{>\kappa})^2-g
\int_{\BbbR^3}dk\frac{\chi^{\Lambda}_{\kappa}(k)}{\sqrt{\omega(k)}}
(a(k)+a(k)^*)
+\dG(\omega^{>\kappa})
-E_{\kappa}^{\Lambda}
\end{align}
and 
\begin{align}
(P-\Pf^{\le \kappa}) \cdot \Pf^{>\kappa}=\sum_{j=1}^3 (P_j-P_{\mathrm{f}, j}^{\le \kappa}) \otimes P_{\mathrm{f}, j}^{>\kappa}.
\end{align}

\subsection{$e^{-\beta \Hr(P)}$ is positivity preserving w.r.t. $\Fock_+$ }

In this subsection, we will show the following proposition.

\begin{Prop}\label{GlobalPP}
For all $ P\in \BbbR^3 $ and $\beta\ge 0$,  we have 
$e^{-\beta \Hr(P)}\unrhd 0$ w.r.t. $\mathfrak{F}_{ +}$.
\end{Prop}

\subsubsection{Proof of Proposition \ref{GlobalPP}}

\begin{lemm}\label{BasicPP1}
We have the following:
\begin{itemize}
\item[(i)] $e^{-\beta \dG(\omega)} \unrhd 0$ w.r.t. $\Fock_+$ for all $\beta \ge 0$.
\item[(ii)] $e^{-\beta (P-\Pf)^2/2} \unrhd 0$ w.r.t. $\Fock_+$ for all $P\in \BbbR^3$ and $\beta \ge 0$.
\end{itemize}
\end{lemm}
{\it Proof.} (i) Note  that $e^{-\beta \omega} \unrhd 0$ w.r.t. $L^2(\BbbR^3)_+$ for all $\beta \ge 0$.
By Proposition \ref{PPFock3}, we obtain (i).

(ii) Note  that 
\begin{align}
e^{-\beta (P-\Pf)^2/2}=\Sumoplus e^{-\beta(P-k_1-\cdots-k_n)^2/2}.
\end{align}
Since each multiplication operator 
$
e^{-\beta(P-k_1-\cdots-k_n)^2/2}
$
preserves the positivity w.r.t. $L^2_{\mathrm{sym}} (\BbbR^{3n})_+$, we conclude (ii). $\Box$
\medskip\\

\begin{lemm}\label{CPP}
$e^{-\beta \Hrl(P)} \unrhd 0$ w.r.t. $\Fock_+$ for all $P\in \BbbR^3$, $\beta \ge 0$ and $\Lambda>0$.
\end{lemm}
{\it Proof.} By Proposition \ref{PPFockII}
 and Lemma \ref{BasicPP1}, we can apply Proposition \ref{BasicPertPP}
 with $A=\frac{1}{2}(P-\Pf)^2+\dG(\omega)$ and $B=-\{a(f)+a(f)^*\},\ \ f=g\frac{\chi_{\Lambda}}{\sqrt{\omega}}$.
 $\Box$
 \medskip
 \begin{flushleft}
{\it 
Proof of Proposition \ref{GlobalPP}
}
\end{flushleft}
Because $e^{-\beta \Hrl(P)}$ strongly converges to $e^{-\beta \Hr(P)}$, the assertion follows from 
Proposition \ref{WeakCl} and  Lemma \ref{CPP}. $\Box$

\subsection{$e^{-\beta \Hr^{\le \kappa}(P)}$ is positivity improving w.r.t. $\Fock_+^{\le \kappa}$}
Our goal here is to prove the following.

\begin{Prop}\label{LocalPI}
For all $P\in \BbbR^3, \beta >0$ and $\kappa>0$, we have 
$
e^{-\beta \Hr^{\le \kappa}(P)} \rhd 0
$ w.r.t. $\Fock_+^{\le \kappa}$.
\end{Prop}

\subsubsection{Proof of Proposition \ref{LocalPI}}

Using arguments similar to those in the proof of Lemma \ref{BasicPP1}, we have the following.

\begin{lemm}\label{BasicRP1}
We have the following:
\begin{itemize}
\item[(i)] $e^{-\beta \dG(\omega^{\le \kappa })} \unrhd 0$ w.r.t. $\Fock_+^{\le \kappa}$ for all $\beta \ge 0$.
\item[(ii)] $e^{-\beta (P-\Pf^{\le \kappa})^2/2} \unrhd 0$ w.r.t. $\Fock_+^{\le \kappa}$ for all $P\in \BbbR^3$ and $\beta \ge 0$.
\end{itemize}
\end{lemm}
\begin{lemm}\label{LocalPP}
For all $P\in \BbbR^3, \beta >0$ and $\kappa>0$, we have 
$
e^{-\beta \Hr^{\le \kappa}(P) }\unrhd 0
$ w.r.t. $\Fock_+^{\le \kappa}$.
\end{lemm}
{\it Proof.} By Proposition \ref{LocalPropErgo} and Lemma \ref{BasicRP1}, we can apply Proposition \ref{BasicPertPP} with  $A=\frac{1}{2}(P-\Pf^{\le \kappa})^2+\dG(\omega^{\le \kappa})$ and 
$B=-\{a(F)+a(F)^*\},\ \ F=g\frac{\chi_{\kappa}}{\sqrt{\omega}}$. $\Box$

\begin{flushleft}
{\it 
Proof of Proposition \ref{LocalPI}
}
\end{flushleft}
Let $\displaystyle F=g\frac{\chi_{\kappa}}{\sqrt{\omega}}$. Because $F(k)>0$ on $B_{\kappa}$,
$\phi(F)=a(F)+a(F)^*$ is ergodic w.r.t. $\Fock_+^{\le \kappa}$ by Proposition \ref{LocalPropErgo}.
Let $\vphi, \psi\in \Fock_+^{\le \kappa} \backslash \{0\}$. We can express $\vphi$ and $\psi$ as 
$
\vphi=\Sumoplus \vphi_n
$ 
and $\psi=\Sumoplus \psi_n$.
Since $\vphi$ and $\psi$ are non-zero, there exist $n_1, n_2\in \BbbN_0$ such that 
$\vphi_{n_1}\neq  0$ and $\psi_{n_2} \neq 0$. By the identifications similar to  (\ref{FinIdn}) and the ergodicity of $\phi(F)$,
 there exists an $\ell\in \BbbN_0$ such that 
 \begin{align}
 \la \vphi_{n_1}|\phi(F)^{\ell} \psi_{n_2}\ra >0. \label{ErgP}
 \end{align}
Since $\vphi\ge  \vphi_{n_1}$ and $\psi\ge \psi_{n_2}$ w.r.t. $\Fock_+^{\le \kappa}$, we have
\begin{align}
\la \vphi|e^{-\beta \Hr^{\le \kappa}(P)} \psi\ra \ge \la \vphi_{n_1}|e^{-\beta \Hr^{\le \kappa}(P)} \psi_{n_2}\ra
\label{Red1}
\end{align}
for all $\beta \ge 0$, by Lemma \ref{LocalPP}.
Let $H_0=\frac{1}{2}(P-\Pf^{\le \kappa})^2+\dG(\omega^{\le \kappa})$.
By the Duhamel formula, we obtain
\begin{align}
e^{-\beta \Hr^{\le \kappa}(P)} =\sum_{j=0}^{\ell}D_j+R_{\ell}\ \ \ \mbox{on $\Ffin^{\le \kappa}$},
\end{align}
where 
\begin{align}
D_j=&\int_0^{t}\dm s_1 \int_0^{t-s_1}\dm s_2 \cdots
 \int_0^{t-\sum_{i=1}^{j-1}s_i}\dm s_j\times 
\no &\times e^{-s_1 H_0}
 \phi(F)e^{-s_2 H_0}\cdots e^{- s_j
 H_0} \phi(F)e^{-(t-\sum_{i=1}^{j}s_i)H_0},\\
R_{\ell}=&\int_0^{t}\dm s_1 \int_0^{t-s_1}\dm s_2 \cdots
 \int_0^{t-\sum_{i=1}^{\ell}s_i}\dm s_{\ell+1} \times \no &
 \times e^{-s_1 H_0}
 \phi(F) e^{-s_2 H_0}\cdots e^{- s_{\ell}
 H_0} \phi(F) e^{-(t-\sum_{i=1}^{\ell+1}s_i)\Hr^{\le \kappa}(P)}.
\end{align} 
Because $e^{-s H_0} \unrhd 0$ and $\phi(F) \unrhd 0$ w.r.t. $\Fock^{\le \kappa}_+$, we know that $\la \vphi_{n_1}|D_j\psi_{n_2}\ra\ge 0$. Similarly, by Lemma \ref{LocalPP}, we have $\la \vphi_{n_1}|R_{\ell} \psi_{n_2} \ra \ge 0$. 
Hence, \begin{align}
\la\vphi_{n_1}|e^{-\beta \Hr^{\le \kappa}(P)} \psi_{n_2}\ra \ge \la \vphi_{n_1}|D_{\ell} \psi_{n_2}\ra. \label{Red2}
\end{align}
Let $G(s_1, \dots, s_{\ell})=\la \vphi_{n_1}| e^{-s_1 H_0}
 \phi(F)e^{-s_2 H_0}\cdots e^{- s_{\ell}
 H_0} \phi(F)e^{-(t-\sum_{i=1}^{\ell}s_i)H_0} \psi_{n_2}\ra$.
 By (\ref{ErgP}), we see that $G(0, \dots, 0)>0$.
Because $G(s_1, \dots, s_{\ell})$ is  positive and continuous, 
we have 
\begin{align}
\la \vphi_{n_1}|D_{\ell} \psi_{n_2}\ra=\int_0^{t}\dm s_1 \int_0^{t-s_1}\dm s_2 \cdots
 \int_0^{t-\sum_{i=1}^{\ell-1}s_i}\dm s_{\ell} G(s_1, \dots, s_{\ell})>0.\label{IntPI}
\end{align}
Combining (\ref{Red1}), (\ref{Red2}) and (\ref{IntPI}), we arrive at $\la \vphi|e^{-\beta \Hr^{\le \kappa}(P)} \psi\ra>0$ for all $\beta >0$. $\Box$

\subsection{Basic properties of $K_{\kappa,  \Lambda}$}
In this subsection, we will show the following.
\begin{Prop}\label{KExt}
For all $\kappa>0$, there exists a self-adjoint operator $K_{\kappa}$ bounded from below such that 
\begin{itemize}
\item[(i)] $e^{-\beta K_{\kappa, \Lambda}}$ strongly converges to $e^{-\beta K_{\kappa}}$ for all $\beta \ge 0$, as $\Lambda\to \infty$;
\item[(ii)] $e^{-\beta K_{\kappa}} \unrhd 0$ w.r.t. $\Fock_+^{>\kappa}$ for all $\beta \ge 0$.
\end{itemize}
\end{Prop}

\subsubsection{Proof of  Proposition \ref{KExt} (i)}
We will apply Nelson's idea \cite{Nelson}.
Choose $K$ such that $\kappa<K<\Lambda$. Let
\begin{align}
\beta(k)=g\frac{1-\chi_{K}(k)}{\omega(k)^{1/2}\{
\omega(k)+k^2/2
\}}.
 \end{align} 
We define an anti-self-adjoint operator $T$ by 
\begin{align}
T=\{
a(G)-a(G)^*\}^-,\ \ \ G=\beta \chi_{\kappa}^{\Lambda}.
\end{align}
The unitary operator $e^T$ is called the {\it Gross transformation}, which was introduced in \cite{EPGross}.
We can check the following (For notational simplicity, we give  somewhat formal  expressions here.):
\begin{itemize}
\item $
e^T \Pf^{>\kappa}e^{-T}=\Pf^{>\kappa}+A+A^*
$, where $A=(A_1, A_2, A_3)$ with $A_j=a(k_jG)$.
\item $e^T a(k)e^{-T}=a(k)+G(k)$.  
\end{itemize}
Let $\tilde{K}_{\kappa, \Lambda}=e^{T}K_{\kappa, \Lambda}e^{-T}$.
Using the above  facts, we obtain the following:
\begin{align}
\tilde{K}_{\kappa, \Lambda}
=&\frac{1}{2} (\Pf^{>\kappa})^2+\Pf^{>\kappa}\cdot A+A^*\cdot\Pf^{>\kappa}
+\frac{1}{2}A^2+\frac{1}{2}A^{*2}+A^*\cdot A\no
&+H_I+\dG(\omega^{>\kappa})-E^{K}_{\kappa} ,   
\end{align}
where 
\begin{align}
H_I=-g\int_{\BbbR^3}dk \frac{\chi_{\kappa}^K(k)}{\sqrt{\omega(k)}}(a(k)+a(k)^*).
\end{align}

We set 
\begin{align}
\mathcal{J}=\frac{1}{2} (\Pf^{>\kappa})^2+\dG(\omega^{>\kappa}).
 \end{align} 
Let us define a quadratic form $B_{\Lambda}$ on $\D(\mathcal{J}^{1/2})\times \D(\mathcal{J}^{1/2})$ by 
\begin{align}
B_{\Lambda}(\vphi, \psi)
=&\sum_{j=1}^3
\Big\{
\la P_{\mathrm{f}, j}^{>\kappa} \vphi|A_j \psi\ra+\la A_j\vphi|P_{\mathrm{f}, j}^{>\kappa} \psi\ra
+\frac{1}{2}\la A_j^*\vphi|A_j\psi\ra+\frac{1}{2}\la A_j\vphi|A_j^*\psi\ra\no
&+\la A_j\vphi|A_j\psi\ra
\Big\}+\la \vphi|H_I\psi\ra. \label{BLD}
\end{align}
We easily check that 
\begin{align}
\la \vphi|\tilde{K}_{\kappa, \Lambda} \psi\ra= \la \mathcal{J}^{1/2} \vphi|\mathcal{J}^{1/2} \psi\ra+B_{\Lambda}(\vphi, \psi),\ \ \vphi, \psi\in \D(\mathcal{J}^{1/2}).
\end{align}

Let $G_{\infty}=\beta (1-\chi_{\kappa})$ and let $A_{\infty}=a(kG_{\infty})$.
We define a quadratic form $B_{\infty}$ on $\D(\mathcal{J}^{1/2}) \times \D(\mathcal{J}^{1/2})$
 by replacing $A$ with $A_{\infty}$ in (\ref{BLD}).
\begin{lemm}\label{BInfEst}
Let $C(K)$ be a positive number defined by  
\begin{align}
C(K)^2 =\int_{\BbbR^3} dk \frac{1-\chi_K(k)}{\{\omega(k)+k^2/2\}^2}. \label{DefCK}
\end{align}
For all $\vepsilon >0$, there exists a constant $D_{K, \vepsilon}>0$ such that  
\begin{align}
|B_{\infty}(\vphi, \vphi)| \le \{
6C(K)+6C(K)^2+\vepsilon
\}\|(\mathcal{J}+1)^{1/2} \vphi\|^2+D_{K, \vepsilon}
\|\vphi\|^2 \label{BInq}
\end{align}
for all $\vphi\in \D(\mathcal{J}^{1/2})$.
\end{lemm}
{\it Proof.} Using (\ref{CreAnnInq}) and (\ref{CreAnnInq2}), we have 
$
\|A^{\#}_{\infty, j} \vphi\| \le \|\omega^{-1/2} k_j G\|\|(\mathcal{J}+1)^{1/2} \vphi\|
$, where $a^{\#} =a$ or $a^*$. Because $
\|\omega^{-1/2} k_j G\| \le C(K)
$, we obtain
\begin{align}
 \|A^{\#}_{\infty, j} \vphi\| \le C(K)\| (\mathcal{J}+1)^{1/2} \vphi\|,\ \ \vphi\in \D(\mathcal{J}^{1/2}). \label{A1}
\end{align}
On the other hand, we have 
\begin{align}
\|P_{\mathrm{f}, j}^{> \kappa} \vphi\| \le \|(\mathcal{J}+1)^{1/2} \vphi\|,\ \ \vphi\in \D(\mathcal{J}^{1/2}).\label{P1}
\end{align}
By using (\ref{A1}) and (\ref{P1}), we can 
estimate the terms involving $A$ and $\Pf^{>\kappa}$.

In order to estimate $\la \vphi|H_I\psi\ra$, we observe, by   (\ref{CreAnnInq}) and (\ref{CreAnnInq2}) again,
\begin{align}
|\la \vphi|H_I\vphi\ra| \le D \|\vphi\| \|(\mathcal{J}+1)^{1/2} \vphi\|,
\end{align}
where $\displaystyle 
D=2g \bigg(
\int dk \frac{\chi_{\kappa}^K}{\omega^2}
\bigg)^{1/2}$. Using $ab \le \vepsilon a^2+b^2/4\vepsilon$, we obtain
\begin{align}
|\la \vphi|H_I\vphi\ra| \le \vepsilon  \|(\mathcal{J}+1)^{1/2} \vphi\|^2+\frac{D}{4\vepsilon} \|\vphi\|^2.
\end{align}
Thus we are done.
$\Box$
\medskip\\

Choose $K$ sufficiently large as $6C(K)+6C(K)^2<1$. By the KLMN theorem \cite[Theorem X. 17]{ReSi2} and Lemma \ref{BInfEst}, there exists a unique self-adjoint operator $\tilde{K}_{\kappa}$ such that 
\begin{align}
\la \vphi|\tilde{K}_{\kappa} \psi\ra= \la \mathcal{J}^{1/2} \vphi|\mathcal{J}^{1/2} \psi\ra+B_{\infty}(\vphi, \psi).
\end{align}
Note  that $\tilde{K}_{\kappa}$ is bounded from below.

\begin{lemm}\label{DiffBL}
We have 
\begin{align}
|B_{\infty}(\vphi, \vphi)-B_{\Lambda}(\vphi, \vphi)|
\le 
\Big\{
6C(\Lambda)+12C(K) C(\Lambda)
\Big\} \|(\mathcal{J}+1)^{1/2} \vphi\|^2 \label{DiffB}
\end{align}
for all $\vphi\in \D(\mathcal{J}^{1/2})$, where $C(K)$ and $C(\Lambda)$ are defiend by (\ref{DefCK}).
\end{lemm}
{\it Proof}.  By (\ref{CreAnnInq}) and (\ref{CreAnnInq2}), we have
\begin{align}
\|(A_{\infty, j}^{\#}-A_j^{\#}) \vphi\|& \le 
\|\omega^{-1/2} k_j \beta (1-\chi_{\kappa} -\chi_{\kappa}^{\Lambda})\|
\|(\mathcal{J}+1)^{1/2} \vphi\|\no
& \le C(\Lambda) \|(\mathcal{J}+1)^{1/2} \vphi\|,\ \ \vphi\in \D(\mathcal{J}^{1/2}). \label{DifferenceA}
\end{align}
Using (\ref{A1}), (\ref{P1}) and (\ref{DifferenceA}), we can prove (\ref{DiffB}). $\Box$

\begin{flushleft}
{\it
Proof of Theorem \ref{KExt} (i)
}
\end{flushleft}
Note  that $C(\Lambda) \to 0$ as $\Lambda\to \infty$.
By Lemma \ref{DiffBL} and \cite[Theorem VIII. 25]{ReSi1}, $\tilde{K}_{\kappa, \Lambda}$ converges to $\tilde{K}_{\kappa}$
 in  norm resolvent sense as $\Lambda\to \infty$. 
 Let $T_{\infty}=\{a(G_{\infty})-a(G_{\infty})^*\}^-$.
 Because $e^T$ strongly  converges to $e^{T_{\infty}}$, we obtain the desired result. $\Box$

\subsubsection{Proof of Proposition \ref{KExt} (ii)}
Using arguments similar to those in the proof of Lemmas \ref{BasicPP1} and \ref{CPP},
we can show the following lemma.

\begin{lemm}\label{KPP}
$e^{-\beta K_{\kappa, \Lambda}} \unrhd 0$ w.r.t. $\Fock_+^{>\kappa}$ for all $\beta \ge 0,\ \kappa>0$ and $\Lambda>0$.
\end{lemm}

\begin{flushleft}
{\it 
Proof of  Proposition \ref{KExt} (ii)
}
\end{flushleft}
By Proposition \ref{KExt} (i), $e^{-\beta K_{\kappa, \Lambda}}$ strongly converges to $e^{-\beta K_{\kappa}}$
as $\Lambda\to \infty$. Using Proposition \ref{WeakCl} and Lemma \ref{KPP}, we conclude Proposition \ref{KExt} (ii). $\Box$

\subsection{A key theorem}

Let 
\begin{align}
L_{\kappa}=
H_{\mathrm{ren}}^{\le  \kappa}(P)\otimes\one +\one \otimes   K_{\kappa}.
\end{align}
Our purpose in this subsection is to prove the following theorem.

\begin{Thm}\label{LocalEq}
The following (i) and (ii) are  mutually equivalent:
\begin{itemize}
\item[(i)] $e^{-\beta \Hr(P)} \rhd 0$ w.r.t. $\Fock_+$ for all $\beta >0$.
\item[(ii)] For each $\vphi, \psi\in \Fock_+\backslash \{0\}$, there exist $\beta\ge 0$ and $\kappa>0$ such that $\la \vphi|e^{-\beta L_{\kappa}} \psi\ra>0$.
\end{itemize}
\end{Thm}

\subsubsection{Proof of Theorem \ref{LocalEq}}
Let $A$ and $B$ be self-adjoint operators, and let $E_A$ and $E_B$ be their spectral measures.
Assume that $E_A$ and $E_B$ commute with each other: $E_A(I) E_B(J)=E_B(J)E_A(I)$ for all $I, J\in \mathbb{B}^1$, the Borel sets of $\BbbR$.
We can decompose $A$ as $A=A_+-A_-$, where $A_+$ and $A_-$ are positive and negative parts of $A$, respectively. Similarly, we have $B=B_+-B_-$. 

For each $n\in \BbbN$, we set 
\begin{align}
(AB)_{[n]} =& A_+B_++A_-B_--\Big(
A_+E_A[0, n] B_- E_B[-n, 0]
+A_-E_A[-n, 0]B_+E_B[0, n]
\Big).
\end{align}
Note  that  $E_A[-n, 0]=E_{A_-} [0, n]$ and $E_B[-n, 0]=E_{B_-}[0, n]$.
Thus, we have 
\begin{align}
(AB)_{[n]} &\ge -2n^2, \label{BddAB1}\\
(AB)_{[n]} &\ge (AB)_{[n+1]}. \label{MonoAB}
\end{align}
Similarly, we define 
\begin{align}
(AB)^{[n]}=A_+E_A[0, n] B_+ E_B[0, n]+A_- E_A[-n, 0] B_- E_B[-n, 0]
-(A_+B_-+A_-B_+).
\end{align}
We have 
\begin{align}
(AB)^{[n]} &\le 2n^2, \label{BddAB2}\\
(AB)^{[n]} &\le (AB)^{[n+1]}. \label{MonoAB2}
\end{align}

For each $\kappa>0$, we define a sequence of  self-adjoint operators $\{C_{\kappa, n}^{+}\}_{n=1}^{\infty}$ by 
\begin{align}
C_{\kappa, n}^+=-\sum_{j=1}^3\Big((P_j-P_{\mathrm{f}, j}^{\le \kappa}) P_{\mathrm{f}, j}^{<\kappa}\Big)_{[n]}.
\end{align}
Similarly, we define
\begin{align}
C_{\kappa, n}^-=-\sum_{j=1}^3\Big((P_j-P_{\mathrm{f}, j}^{\le \kappa}) P_{\mathrm{f}, j}^{<\kappa}\Big)^{[n]}.
\end{align}
Let $C_{\kappa}=-(P-\Pf^{\le \kappa}) \cdot \Pf^{>\kappa}$.
Note  that $C_{\kappa, n}^{\pm} \vphi$ converges to $C_{\kappa}\vphi$  as $n\to \infty$ for each $\vphi\in \D(C_{\kappa})$.
By (\ref{BddAB1}), (\ref{MonoAB}), (\ref{BddAB2}) and (\ref{MonoAB2}), we have
\begin{align}
C_{\kappa, n}^+&\le 6n^2, \label{C1}\\
C_{\kappa, n}^+&\le C_{\kappa, n+1}^+, \label{C2}\\
C_{\kappa, n}^- &\ge -6n^2, \label{C3}\\
C_{\kappa, n}^-&\ge C_{\kappa, n+1}^-. \label{C4}
\end{align}

\begin{lemm}\label{PPC}
For all  $n\in \BbbN$ and $s\ge 0$, we have the following:
\begin{itemize}
\item[(i)]
$e^{-sC_{\kappa, n}^-}
$ is bounded and 
$e^{-sC_{\kappa, n}^-} \unrhd 0$ w.r.t. $\Fock_+$.
\item[(ii)]
$e^{sC_{\kappa, n}^+}
$ is bounded and 
$e^{sC_{\kappa, n}^+} \unrhd 0$ w.r.t. $\Fock_+$.
\end{itemize}
\end{lemm}
{\it Proof.} 
(i) By (\ref{C3}), $e^{-s C_{\kappa, n}^-}$ is bounded for all $s\ge 0$.
We can express $e^{-sC_{\kappa, n}^-}$ as 
\begin{align}
e^{-s C_{\kappa, n}^-}=\sideset{}{^{\oplus}_{\ell\ge 0}}\sum F_{\ell}, \label{DirectC}
\end{align}
where $F_{\ell}$ is some multiplication operator on $L^2_{\mathrm{sym}}(\BbbR^{3\ell})$.
We easily see that the function $F_{\ell}$ is positive. Thus, $F_{\ell}\unrhd 0$ w.r.t. $L^2_{\mathrm{sym}}(\BbbR^{3\ell})_+$ for all $\ell\in \BbbN$, which implies (i). Similarly, we can prove (ii). 
$\Box$

\begin{lemm}\label{PPVan}
Let $\vphi, \psi\in \Fock_+$. 
\begin{itemize}

\item[(i)]If $\la \vphi|\psi\ra=0$, then
$\la \vphi|e^{-sC_{\kappa, n}^-}\psi\ra=0$   for all $n\in \BbbN,\ s\ge 0$ and $\kappa>0$.
\item[(ii)]If $\la \vphi|\psi\ra=0$, then
$\la \vphi|e^{sC_{\kappa, n}^+}\psi\ra=0$   for all $n\in \BbbN,\ s\ge 0$ and $\kappa>0$.
\end{itemize}
\end{lemm}
{\it Proof.} (i) 
We can express $\vphi$ and $\psi$ as 
\begin{align}
\vphi=\sideset{}{^{\oplus}_{\ell \ge 0}}\sum \vphi_{\ell},\ \ \ \psi=\sideset{}{^{\oplus}_{\ell\ge 0}}\sum \psi_{\ell}.
\end{align}
Note  that 
$\vphi_{\ell}$ and $\psi_{\ell}$ are positive functions  in $L_{\mathrm{sym}}^2(\BbbR^{3\ell})$.
The condition $\la \vphi|\psi\ra=0$ is equivalent to the condition $\la \vphi_{\ell}|\psi_{\ell}\ra=0$
 for all $\ell\in \BbbN_0$.
Recall the expression (\ref{DirectC}). Because $F_{\ell}$ is positive and bounded, we conclude that 
$\la \vphi_{\ell}|F_{\ell}\psi_{\ell}\ra=0$, which implies that 
$
\la \vphi|e^{-sC_{\kappa, n}^-} \psi\ra=\sum_{\ell=0}^{\infty} \la \vphi_{\ell}|F_{\ell}\psi_{\ell}\ra=0.
$
Similarly, we can prove (ii). $\Box$

\begin{lemm}\label{LKPP}
$e^{-\beta L_{\kappa}} \unrhd 0$ w.r.t. $\Fock_+$ for all $P\in \BbbR^3$ and $\beta \ge 0$.
\end{lemm}
{\it Proof.} By Propositions \ref{TensPPAB}, \ref{LocalPI} and  \ref{KExt}, we obtain the assertion in the lemma. $\Box$

\begin{lemm}\label{StReC}
We  have the following:
\begin{itemize}
\item[(i)] $L_{\kappa}\dot{+}C_{\kappa, n}^-$ converges to $\Hr(P)$ in strong resolvent sense as $n\to \infty$, where $\dot{+}$ in dicates the form sum.
\item[(ii)] $\Hr(P)\dot{-}C_{\kappa, n}^+$ converges to $L_{\kappa}$ in strong resolvent sense as $n\to \infty$.
\end{itemize}
\end{lemm}
{\it Proof.} (i)
Let us define a sequence of closed, positive quadratic form  $\{t_n\}_{n=1}^{\infty}$ by 
\begin{align}
t_n(\vphi, \psi)=\la \vphi| \{L_{\kappa}+C_{\kappa, n}^-+ Const.\}\psi\ra,
\end{align}
where $Const.$ is chosen such that $t_n$ is uniformly positive.
By (\ref{C4}), we have $t_1\ge t_2\ge \cdots\ge t_n \ge \cdots$
and $\lim_{n\to \infty}t_n(\vphi, \vphi)=t_{\infty}(\vphi, \vphi)$, where $t_{\infty}$ is a quadratic form associated with $\Hr(P)$.
Thus, by \cite[Theorem S. 16]{ReSi1}, we obtain (i).

Similarly, we can prove (ii)
 by applying \cite[Theorem S. 16]{ReSi1}. $\Box$
 \medskip

\begin{flushleft}
{\it 
Proof of Theorem \ref{LocalEq}
}
\end{flushleft}
(i) $\Longrightarrow$ (ii):
We extend the idea  in \cite{Faris}.
Let $\psi\in \Fock_+\backslash \{0\}$. We set 
$
K(\psi)=\{
\vphi\in \Fock_+\, |\, \la \vphi| e^{-\beta L_{\kappa}} \psi\ra=0\ \forall \beta \ge 0\ \forall \kappa>0
\}
$.
We will show that $K(\psi)=\{0\}$.
Let $\vphi\in K(\psi)$: $\la \vphi|e^{-\beta L_{\kappa}} \psi\ra=0$
for all $\beta \ge 0$ and $\kappa>0$.
By Lemma \ref{PPVan} (i) and Lemma \ref{LKPP}, we have $\la e^{-s C_{\kappa, n}^-} \vphi|e^{-\beta L_{\kappa}} \psi\ra=0$
for all $n\in \BbbN,\ s\ge0,\ \beta \ge 0$ and $\kappa>0$, 
which implies that $e^{-sC_{\kappa, n}^-} K(\psi)\subseteq K(\psi) $.
On the other hand, it is easy to check that $e^{-t L_{\kappa}} K(\psi)\subseteq K(\psi)$
for all $t\ge 0$.
Hence, $
(e^{-\beta L_{\kappa}/\ell} e^{-\beta C_{\kappa, n}^-/\ell})^{\ell} K(\psi)\subseteq K(\psi)
$ for all $\ell\in \BbbN$. Taking $\ell\to \infty$, we obtain that $
e^{-\beta (L_{\kappa}\dot{+}C_{\kappa, n}^-)} K(\psi)\subseteq K(\psi)
$ for all $n\in \BbbN$ and $\beta \ge 0$ by \cite[Theorem S. 21]{ReSi1}.
Taking $n\to \infty$, we arrive at $
e^{-\beta \Hr(P)} K(\psi) \subseteq K(\psi)
$ for all $\beta \ge 0$ by Lemma \ref{StReC} (i). Therefore, for each $\vphi\in K(\psi)$, it holds that 
$\la \vphi|e^{-\beta\Hr(P)} \psi\ra=0$ for all $\beta \ge 0$. By the assumption (i),
$\vphi$ must be $0$.
\medskip

(ii) $\Longrightarrow$ (i):
We will provide a sketch. For each $\psi\in \Fock_+\backslash \{0\}$, we set 
$
J(\psi)=\{
\vphi\in \Fock_+\, |\, \la \vphi| e^{-\beta \Hr(P)} \psi\ra=0\ \forall \beta \ge 0
\}
$. Using arguments similar to those in the previous part, we can show that $
e^{-\beta L_{\kappa}} J(\psi) \subseteq J(\psi)
$ for all $\beta \ge 0$ and $\kappa>0$. Thus, for each $\vphi\in J(\psi)$, we obtain 
$\la \vphi|e^{-\beta L_{\kappa}} \psi\ra=0$ for all $\beta \ge 0$ and $\kappa>0$. 
By the assumption (ii), $\vphi$ must be $0$, which implies $J(\psi)=\{0\}$. 
Thus, for each $\vphi, \psi\in \Fock_+\backslash \{0\}$, there exists a $\beta \ge 0$ such that 
$\la \vphi| e^{-\beta \Hr(P)} \psi\ra >0$. Applying Theorem \ref{PFF}, we conclude (i).
$\Box$

\subsection{Completion of proof of Theorem \ref{Main1}}

\begin{Prop}\label{KeyInq}
For all $P\in \BbbR^3$ and $ \kappa>0$, we have
\begin{align}
e^{-\beta L_{\kappa}}  \unrhd \la \Omega^{>\kappa}|e^{-\beta K_{\kappa}} \Omega^{>\kappa} \ra e^{-\beta \Hr^{\le \kappa}(P)} \otimes \one Q_{\kappa}
\end{align}
w.r.t. $\Fock_+$, where $\Omega^{>\kappa}$ is the Fock vacuum in $\Fock^{>\kappa}$.
\end{Prop}
{\it Proof.} By Proposition \ref{QPP}, it holds that $Q_{\kappa} \unrhd 0$ and $Q_{\kappa}^{\perp} \unrhd 0$ w.r.t. $\Fock_+$.
Thus,  by Lemma \ref{LKPP},
\begin{align}
e^{-\beta L_{\kappa}} \unrhd Q_{\kappa} e^{-\beta L_{\kappa}} Q_{\kappa}\ \ \mbox{w.r.t. $\Fock_+$ for all $\beta \ge 0$}.
\end{align}
By Lemma \ref{QAct}, we have
\begin{align}
\la \vphi|Q_{\kappa}e^{-\beta L_{\kappa}}Q_{\kappa} \psi\ra
=&\la \vphi_{\kappa}\otimes \Omega^{>\kappa}|e^{-\beta L_{\kappa}} \psi_{\kappa}\otimes \Omega^{>\kappa}\ra\no
=& \la \Omega^{>\kappa} |e^{-\beta K_{\kappa}} \Omega^{>\kappa}\ra\la \vphi_{\kappa}|e^{-\beta \Hr^{\le \kappa}(P)} \psi_{\kappa}\ra\no
=&\la \Omega^{>\kappa} |e^{-\beta K_{\kappa}} \Omega^{>\kappa}\ra\la \vphi |e^{-\beta \Hr^{\le \kappa}(P)} \otimes \one Q_{\kappa}\psi\ra,
\end{align}
which implies that 
$
Q_{\kappa}e^{-\beta L_{\kappa}}Q_{\kappa} 
=\la \Omega^{>\kappa}|e^{-\beta K_{\kappa}} \Omega^{>\kappa} \ra e^{-\beta \Hr^{\le \kappa}(P)} \otimes \one Q_{\kappa}
$.
Here, we used the fact that 
$Q_{\kappa} e^{-\beta \Hr^{\le \kappa}(P)} \otimes 1 Q_{\kappa}=e^{-\beta \Hr^{\le \kappa}(P)} \otimes 1 Q_{\kappa}$,  which follows from Proposition \ref{CommuL}. 
  $\Box$

\begin{lemm}\label{Nenno}
$
\la \Omega^{>\kappa}|e^{-\beta K_{\kappa}} \Omega^{>\kappa}\ra>0
$ for all $\beta\ge 0$ and $\kappa>0$.
\end{lemm}
{\it Proof.} Because $\ker(e^{-\beta K_{\kappa}})=\{0\}$ by Proposition \ref{KExt}, the assertion is easy to check. $\Box$

\begin{flushleft}
{\it Proof of Theorem \ref{Main1}
}
\end{flushleft}
Let $\vphi, \psi\in \Fock_+\backslash \{0\}$. Because $Q_{\kappa}$ strongly converges to $1$ as $\kappa\to \infty$, there exists a $\kappa>0$ such that $Q_{\kappa} \vphi \neq 0$ and $Q_{\kappa} \psi \neq 0$.
By Proposition \ref{KeyInq}, we have
\begin{align}
\la \vphi|e^{-\beta L_{\kappa}} \psi\ra \ge 
\la\Omega^{>\kappa}|e^{-\beta K_{\kappa}} \Omega^{>\kappa} \ra
\la \vphi| e^{-\beta \Hr^{\le \kappa}(P)} \otimes \one Q_{\kappa} \psi\ra.\label{EqLQ2}
\end{align}
Remark that 
\begin{align}
\la \vphi| e^{-\beta \Hr^{\le \kappa}(P)} \otimes \one Q_{\kappa} \psi\ra
=\la \vphi_{\kappa}|e^{-\beta \Hr^{\le \kappa}(P)} \psi_{\kappa}\ra, \label{EqLQ}
\end{align}
 where $\vphi_{\kappa}$ and $\psi_{\kappa}$ are defined  by (\ref{Defpsik}).
 Of course, $\vphi_{\kappa} \neq 0$ and $\psi_{\kappa} \neq 0$.
By Proposition \ref{LocalPI}, the right hand side of (\ref{EqLQ}) is strictly positive, provided that $\beta>0$.
Because $
\la\Omega^{>\kappa}|e^{-\beta K_{\kappa}} \Omega^{>\kappa} \ra>0
$ by Lemma \ref{Nenno}, 
we know  that the right hand side of (\ref{EqLQ2}) is strictly positive.
By Theorem \ref{LocalEq}, we finally conclude that $e^{-\beta \Hr(P)} \rhd 0$ w.r.t. $\Fock_+$
 for all $P\in \BbbR^3$ and $\beta>0$.
 $\Box$

\appendix

\section{A useful proposition }\label{AppA}
\setcounter{equation}{0}
In this appendix, we will review a  useful result concerning  the
operator inequalities introduced in Section \ref{SecMono}.

\begin{Prop}\label{BasicPertPP}
Let $A$ be a positive self-adjoint operator and let $B$ be a  symmetric
 operator. Assume the following:
\begin{itemize}
\item[(i)] $B$ is $A$-bounded with relative bound $a<1$, i.e.,
                 $\D(A)\subseteq \D(B)$ and $\|Bx\|\le a \|Ax\|+b\|x\|$
                 for all $x\in \D(A)$.
\item[(ii)] $0\unlhd e^{-tA}$ w.r.t. $\Cone$ for all $t\ge 0$.
\item[(iii)]$0\unlhd -B$ w.r.t. $\Cone$.
\end{itemize} 
Then $0\unlhd e^{-t(A+B)}$ w.r.t. $\Cone$ for all $t\ge 0$.
\end{Prop}
{\it Proof.}
This proposition is already proved in  \cite{Miyao}, see also \cite{ Miyao3,
Miyao4, Miyao5}. For readers' convenience, we provide    a proof.

  For each $\vepsilon >0$, we set $B_{\vepsilon}=e^{-\vepsilon A} B e^{-\vepsilon A}$.
By (i) and (iii), $B_{\vepsilon}$ is bounded and $-B_{\vepsilon} \unrhd 0$ w.r.t. $\Cone$.
Let us consider  a self-adjoint operator $C_{\vepsilon}=A+B_{\vepsilon}$.
By the Duhamel formula, we have the following norm convergent expansion:
\begin{align}
e^{-t C_{\vepsilon}}&=\sum_{n=0}^{\infty}D_n, \label{Duha1}\\
D_n&=\int_{S_n(t)} e^{-s_1 A}(-B_{\vepsilon}) e^{-s_2 A}(-B_{\vepsilon})\cdots
 e^{-s_n A} (-B_{\vepsilon}) e^{-(\beta-\sum_{j=1}^ns_j)A},
\end{align} 
where $\int_{S_n(t)}=\int_0^{\beta}ds_1\int_0^{\beta-s_1}ds_2\cdots
\int_0^{\beta-\sum_{j=1}^{n-1}s_j} ds_n$ and
$D_0=e^{-t A}$. Since $-B_{\vepsilon} \unrhd 0$ and $\ex^{-t
A}\unrhd 0$ w.r.t. $\Cone$ for all $t \ge 0$, it holds that, by Lemma \ref{Miura},
\begin{align}
\underbrace{ 
e^{-s_1 A}
}_{\unrhd 0}
\underbrace{(-B_{\vepsilon})}_{\unrhd 0}
\underbrace{ 
e^{-s_2 A}
}_{\unrhd 0}\cdots
 \underbrace{e^{-s_n A}}_{\unrhd 0} 
\underbrace{(-B_{\vepsilon})}_{\unrhd 0}
\underbrace{e^{-(t-\sum_{j=1}^ns_j)A}}_{\unrhd 0} \unrhd 0,
\end{align} 
provided that $s_1 \ge 0, \dots, s_n\ge 0$ and $t-s_1-\cdots-s_n\ge 0$. Thus, by Proposition \ref{WeakCl},
we obtain
$D_n\unrhd 0$ w.r.t. $\Cone$ for all $n\ge 0$.
 Accordingly,  by (\ref{Duha1}), we have $\ex^{-t{C_{\vepsilon}}}\unrhd
 D_{n=0}=e^{-t A} \unrhd 0$ w.r.t. $\Cone$ for all
 $t \ge 0$ and $\vepsilon \ge 0$. 
 Because $e^{-tC_{\vepsilon}}$ strongly converges to $e^{-t(A+B)}$ as $\vepsilon \to +0$, we conclude that 
 $e^{-t(A+B)} \unrhd 0$ w.r.t. $\Cone$ for all $t\ge 0$ by Proposition \ref{WeakCl}.
 $\Box$

\end{document}